# Miniature Giants: Investigating Limb Long Bone Structure in Dwarf Proboscideans

*by* Camille Bader[1]*, Ursula B. Göhlich[2] and Alexandra Houssaye[1]

[1]Département Adaptations du Vivant, UMR 7179, Mécanismes adaptatifs et Évolution (MECADEV) CNRS/Muséum national d'Histoire naturelle, Paris, France; camille.bader@edu.mnhn.fr, alexandra.houssaye@mnhn.fr

[2]Naturhistorisches Museum Wien, Geological-paleontological Dept., Vienna, Austria; ursula.goehlich@nhm.at

*Corresponding author

# Abstract

Terrestrial vertebrates rely on their skeleton to provide structural support and allow the movement of the body. Heavy, graviportal taxa, such as extant elephants, exhibit numerous adaptive features in their bone anatomy enabling them to withstand their immense weight. Conversely, dwarfing events impose novel biomechanical constraints on the skeleton. The case of dwarf elephants raises the question of how graviportal animals adapt when undergoing drastic size reduction, and whether they retain graviportal features or exhibit paedomorphic traits. In this study, we examine the morphology and microanatomy of the six long bones in two fossil species of dwarf elephants, *Palaeoloxodon tiliensis* (adults) and *P. falconeri* (juveniles), using both quantitative and qualitative approaches (3D geometric morphometrics, virtual slice comparisons, compact bone thickness cartographies). Our results show that in *P. tiliensis*, the reduction in body mass is reflected in the morphology and microanatomy of the bones, suggesting a more flexed limb posture and a reduced parasagittal orientation compared to in fully graviportal proboscideans. Despite this shift, several key graviportal adaptations are retained in *P. tiliensis*. Additionally, comparisons with early, non-graviportal proboscideans indicate that dwarf elephants exhibit a partial reversion to ancestral limb traits, while maintaining essential weight-bearing features. Finally, adult *P. tiliensis* and juvenile *P. falconeri* specimens exhibit a medullary area filled with trabecular bone, similar to that of extant elephants. Dwarf elephants are thus not scaled-down versions of their mainland ancestors, but instead display a combination of juvenile, graviportal and ancestral traits reflecting the impact of dwarfism on their evolutionary trajectory.

# INTRODUCTION

Throughout evolutionary history, many vertebrate lineages have evolved toward gigantism; among them is the proboscidean order, comprising modern elephants and their extinct relatives (Osborn 1936, 1942; Shoshani & Tassy 2005). Proboscideans evolved from relatively small forms in the Eocene to truly gigantic ones from the Miocene onwards, reaching record mass and size among terrestrial mammals, with some species reaching up to nearly twenty tons (Cantalapiedra *et al.* 2021; Larramendi 2016). Some proboscidean lineages, however, experienced the opposite trend during the Pleistocene, undergoing insular dwarfism, a process characterized by a reduction in body size due to limited resources and space in insular environments (Case 1978; Foster 1964). This phenomenon is well-documented in numerous vertebrate taxa, including mammals (e.g. Cretan dwarf hippopotamuses, Sicilian red deer, Balearic Islands cave goats), reptiles (e.g. *Magyarosaurus* dinosaurs, Madagascar dwarf chameleons) and birds (e.g. King Island emus, Hawaiian flightless ibises), and has been observed in various localities around the world (e.g. Valen 1973; Sondaar 1977; Lomolino 1985; Sander *et al.* 2006).

During the Pleistocene, a series of low sea level periods allowed several waves of colonization of the Mediterranean islands by elephantid species, which became isolated with the conclusion of the Ice Age and rising sea levels (Ambrosetti 1968). The following long periods of isolation resulted in the emergence of various species of differing body sizes, sometimes coexisting on the same island (Herridge 2010). Sicily, for instance, witnessed multiple waves of proboscidean colonization (Bertram 2016) and is home to one of the most well-known dwarf proboscidean species, *Palaeoloxodon falconeri* (Busk, 1867) (Palombo 2001). Most of the Mediterranean dwarf "elephants" have been classified within the genus *Palaeoloxodon*, i.e. sister-taxa to the continental straight-tusked elephant *Palaeoloxodon antiquus.* There are at least two exceptions, in which dwarf elephant species are thought to belong to the mammoth line: *Mammuthus lamarmorai* (Palombo 2001), estimated between 420 kg to over 1.5t (Palombo *et al.* 2017), and *Mammuthus creticus*, which was estimated at around 180 kg (Herridge & Lister 2012; Larramendi 2016), making it the smallest mammoth ever found. While common in the Mediterranean islands (e.g. Sicily-Malta, Crete, Tilos), insular dwarfism in proboscideans also occurred in the Channel Islands of California, where an isolated population of Columbian mammoths evolved into the pygmy mammoth, *Mammuthus exilis*. Although much larger than Mediterranean dwarf species, *M. exilis* was still significantly smaller than its mainland counterpart, weighing around two tons, approximately five times less than *M. columbi* (Agenbroad 2012; Larramendi 2016). While essentially concerning elephantid species, dwarfism also occurred in stegodontids, concomitantly with that experienced by elephantids in the Channel Islands and the Mediterranean islands: dwarfed species (approx. ten species depending on attribution; e.g. Aiba *et al.* 2010; van der Geer *et al.* 2016) of stegodontids have been found in several islands of Indonesia and of the Philippines (e.g. *Stegodon sompoensis*, *S. timorensis*, *S. florensis*), the smallest of them, *Stegodon sondaari* on Flores Island, estimated at around 350-400 kg (Larramendi 2016; van den Bergh 1999; van den Bergh *et al.* 2008), five times lighter than the non-dwarf *Stegodon* species *Stegodon zdansky*.

A dwarfing event implies numerous musculoskeletal and physiological changes (Hayashi *et al.* 2023; Sondaar 1977; van der Geer 2005). Biomechanical constraints, in particular, are greatly impacted by a reduction of weight and thus of the mechanical load (Bertram 2016). Due to their prominent role in the support and movement of the body, limb bones are heavily affected by changes in body size and mass and their structure should thus be modified during the process of insular dwarfism. If dwarfism was expressed by isometric scaling of a large body sized species, without alteration of the robustness or the proportions of the limb segments, the species would display a severely over-built body for the reduced loading regime (Bertram 2016), as is observed in dwarf hippos (Georgitsis *et al.* 2022; Prothero & Sereno 1982). This would particularly be true for dwarfism in giant animals such as Pleistocene proboscideans, as these species were among the heaviest mammals that ever existed. Non-dwarf species of graviportal proboscideans such as mastodons, mammoths and gomphotheres, display specific adaptations in their skeletal architecture to accommodate their massive weight (Osborn 1929; Gregory 1912; Coombs 1978; Hutchinson *et al.* 2003; Bader *et al.* 2024). Their appendicular skeleton comprises robust bones (i.e. broader shaft for a given length), a relative stylopod lengthening and autopod shortening, and large feet with shorter phalanges. In addition to these morphological adaptations, their limb long bones also display a microanatomical pattern highly efficient to deal with high body mass in columnar limbs (Nganvongpanit *et al.* 2017; Bader *et al.* 2024). Extant elephants have bones with a relatively high cortical thickness, compared to in other quadrupedal mammals, and their medullary area is almost entirely filled with trabecular bone, allowing for a better transmission of the mechanical load along the bones. In addition, the trabecular tissue is highly anisotropic with trabeculae oriented parasagittally, following the angle of the columnar limb, with bones almost orthogonal to the ground, relying on axial compression to handle the massive body weight. Graviportal proboscideans have evolved such adaptations allowing them to handle increasingly large body sizes. Thus, we expect that in dwarf species, the reduced biomechanical constraints of a much lighter body will be reflected in their limb long bone external morphology and microanatomy. Furthermore, we wonder whether these variations resemble plesiomorphic traits of lighter, non-graviportal, early proboscideans, or if the bone anatomy retains similarity to that of the heavier graviportal elephantids. However, several studies on dwarf elephant remains indicated limited variation in the skeletal morphology compared to non-dwarf species apart from a relative shortening of their autopod bones (Sondaar 1977; Roth 1984; Scarborough 2020). Such limited morphological variation between species of such contrasting sizes is unexpected, so that it has been proposed that the size decrease was instead reflected in their bone microanatomy (Bertram 2016), as shown by the relatively thinner cortex in dwarf elephant tibiae as compared to non-dwarf specimens, although they fell well within the lower range for mammalian long bones, indicating robust, thick-walled long bones when compared to non-proboscidean taxa (Currey 2002; Bertram 2016). However, this assertion is based on tibiae from only two specimens and thus necessitates further analyses.

Dwarfism is sometimes associated with paedomorphosis, a phenomenon characterized by an organism retaining juvenile traits into its adult form (Gould 1977) (e.g. juvenile body proportions in tree kangaroos, (Johnson & Delean 2003); short snout and high metabolic rate in pipistrellus bats, (Genoud & Christe 2011); flatter face in bonobos, (Lieberman *et al.* 2007)). Some studies concluded that dwarf elephants display no paedomorphic growth pattern (Roth 1984; van der Geer *et al.* 2018), while others proposed partial paedomorphic patterns in the shape of the skull (Lister 1989; Palombo 2001) and of the appendicular skeleton (Scarborough 2020). Thus, growth patterns in dwarf elephants are still debated.

In addition, while numerous studies described the anatomy of dwarf elephants' limb bones, they often focused on isolated bones and almost never addressed their microanatomy, with the unique mention of a *Mammuthus exilis* tibia in Curtin *et al.* 2012. Thus, the scientific literature regarding skeletal anatomy in dwarf elephant species lacks comprehensive studies of shape and microanatomical patterns in dwarf elephants' limb long bones.

Juvenile extant elephants exhibit bones that are generally less robust than those of adults, with distinct microanatomical differences such as a relatively thinner cortex and a more limited extension of trabecular bone in the medullary cavity (Bader *et al.* 2025). Examining the shape and microanatomy of limb bones of dwarf species, including juvenile specimens, might provide insights regarding potential paedomorphic patterns associated with insular dwarfism.

Here we study the six limb long bones conjointly, combining the methods and analyses developed in the previous studies on elephant limb long bones (shape and robustness analyses (Bader *et al.* 2023, 2024), microanatomical descriptions (Bader *et al.* 2025)) to study both the inner and outer anatomy of the bones. Together, they will allow for a holistic approach, providing an in-depth understanding of the effect of insular dwarfism on the appendicular skeleton of dwarf elephants. In this study we aim to 1) describe the external morphology and microanatomy of the six long bones in adult dwarf elephants and how they differ from those of non-dwarf elephants, 2) compare the shape, microanatomy and growth patterns of dwarf species to those of juveniles of extant, non-dwarf elephants and 3) compare the external morphology and microanatomy of dwarf species with those of early, relatively small non-columnar proboscideans. This will provide insights on whether dwarf elephants are isometrically scaled-down versions of their mainland ancestors, or if they exhibit a different pattern entirely, either by resembling juveniles of giant proboscidean species or by exhibiting traits in common with early, non-graviportal species; or alternatively, showing unique combinations of these features.

# MATERIAL AND METHOD

## Sample & 3D imaging

Explanations of all institutional abbreviations are provided in Table S1.

For this study we selected 14 bones from two dwarf elephantid species: *Palaeoloxodon tiliensis* (Theodorou, Symeonidis & Stathopoulou, 2007) (8 bones from adult specimens) and *Palaeoloxodon falconeri* (one juvenile femur and five bones from a fetus/neonate specimen). Both taxa are native to Tilos Island and Sicily, respectively, and are thought to have evolved independently from the same non-dwarf, mainland ancestor, *Palaeoloxodon antiquus* (Larramendi & Palombo 2015; Mitsopoulou *et al.* 2015); however, despite both species being dwarf taxa, *P. tiliensis* (approx. 1300 kg, Larramendi 2016) is much larger and heavier than *P. falconeri* (approx. 300 kg, Larramendi 2016).

With the objective of representing the largest diversity of elephantid species and of ontogenetic stages, we added, depending on availability, 398 bones from 161 specimens belonging to 14 elephantid species (Fig. 1A). A large number of these bones were used in previous studies (Bader *et al.* 2023, 2024, 2025). These specimens, which are from several European and American institutions, consist of 78 humeri, 53 radii, 61 ulnae, 85 femora, 75 tibiae and 46 fibulae (Suppl. Table 1). Age determination was sometimes provided by the institutions (Suppl. Table 1). Otherwise, the ontogenetic stage (juvenile, subadult, adult) was assumed based on the level of fusion and development of the epiphyses (juvenile: unfused epiphyses, subadult: visible epiphyseal plate line, adult: fully fused epiphyses).

Among the non-dwarf specimens, 21 bones belong to juvenile specimens of extant elephants (*Elephas maximus*, *Loxodonta africana*, *Loxodonta cyclotis*). Age was sometimes available, otherwise the ontogenetic stage was assumed based on the level of fusion and development of the epiphyses (calf: absent or unfused epiphyses, juvenile: partially fused epiphyses, subadult: fused epiphyses with visible epiphyseal line, adult: fully fused epiphyses; Suppl. Table 1). The sex was generally unknown and could thus not be accounted for in our analyses.

The comparative sample that was used in the geometric morphometrics (GMMs) analyses and microanatomical descriptions is composed of 316 bones, obtained by surfacic methods (n=285) or microtomography (n=29) (see below). The surfacic sample mainly consists of adult elephantid specimens (dwarves and non-dwarves), as well as juveniles of extant elephants to compare the morphological variation among elephantids. Moreover, 29 specimens of adult and juveniles of both dwarf and extant elephants were CT-scanned for a previous study (Bader *et al.* 2025), enabling comparisons of their microanatomy. In order to enhance our comparative sample, we added 84 bones of extant elephants (*E. maximus* and *L. africana*) at varying ontogenetic stages, which had already been scanned by medical tomography, thus providing low resolution scans that could nevertheless be used for a broad visualization of the inner structure for qualitative comparisons and for robustness calculations (Fig. 1B).

## Institutional abbreviations

IMNH, Idaho Museum of Natural History, Pocatello, USA; MNHN, Muséum national d'Histoire naturelle, Paris, France; NHMUK, Natural History Museum, London, UK; NHMW, Naturhistorisches

Museum Wien, Vienna, Austria; NMB, Naturhistorisches Museum Basel, Basel, Switzerland; RVC, Royal Veterinary College, London, UK.

## X-ray microtomographic data

Five bones of *P. falconeri* were scanned at the AST-RX platform (UMS 2700, Muséum National d'Histoire Naturelle, Paris) with reconstructions performed using X-Act RX-Solutions with voxel size varying from 35 µm to 43 µm and 8 bones of *P. tiliensis* were imaged using X-ray micro-computed tomography (microCT, YXLON FF35 CT, Perkin Elmer Y.Panel 4343 CT detector) at the Natural History Museum in Vienna with reconstructions performed using CERA (Siemens), and voxel sizes varying from 48 µm to 83 µm. Moreover, 16 CT-scans were available from for a previous study (Bader *et al.* 2025), including adult and juvenile specimens of extant elephants.

The 84 additional bones were previously scanned using medical computed tomography (Picker PQ5000; Philips Healthcare, Andover, MA, USA) at the Royal Veterinary College's Equine Diagnostic Unit with voxel size varying from 690 µm to 5000 µm.

## Surfacic models for shape analysis

We used 13 bones of the dwarf taxa that were microCT-scanned, since the quality of the scans allowed for precise reconstruction of their surface. Their external surface was segmented and reconstructed in VGStudio MAX (2016, v. 2.2, Volume Graphics Inc.) and each mesh was decimated to reach 250,000 vertices and 500,000 faces using Avizo 8.1 (VSG, Burlington, MA, USA). The surface of the 84 bones that were obtained via medical tomography was segmented using VGStudio MAX (2016) and used in the robustness measurements only, the imprecise quality of surfacic details preventing their use in the shape analyses. Most of the 3D models of the bones from the comparative sample were already available (Bader *et al.* 2023, 2024, 2025).

## Virtual cross-sections

Virtual cross-sections of the CT-scanned bones (n=113) were made with VGStudio Max with each bone oriented in anatomical position following (Smuts & Bezuidenhout 1994, 1993). We defined the coronal and sagittal sections as cutting through the medullary area, and the transverse section as being perpendicular to the longitudinal axis of the bone and cutting through its center of ossification. Due to their morphological specificities (olecranon tuberosity of the ulna and torsion of the fibular diaphysis), additional cross-sections were added for the ulna and fibula, following Bader *et al.* (2025).

## Bone thickness mapping

We created 3D bone thickness maps of all bones scanned by micro-CT (n=29), which allows sufficient resolution to discern between trabecular and cortical tissue. 3D-mapping of the bone cortical thickness provides graphical outputs allowing for qualitative comparisons of the bone cortical distribution. In order to obtain 3D cortical thickness maps, cortical and trabecular bone need to be separated. Following Bader *et al.* (2022), we manually isolated an outer surface (outer surface of the bone) and an inner surface (inner limit of the compact cortex) for each bone, using Avizo 8.1. We then generated 3D bone cartographies using the 'SurfaceDistance' module in Avizo, i.e. calculating the thickness of cortical bone by measuring the distance between the outer and the inner surfaces of the cortex, and generating 3D cortical thickness maps of the entire bones using relative (i.e. normalized to the bone minimum and maximum) values. We used the 3D mappings to compare the patterns of cortical thickness distribution between dwarf and non-dwarf specimens, choosing *L.*

*africana*, rather than *E. maximus*, as the extant elephant reference due to its closer phylogenetic relationship to the genus *Palaeoloxodon*.

# Shape analyses – Geometric morphometrics

## Landmark digitization

We defined the shape of the bones using anatomical landmarks and curve sliding semi-landmarks, as described by Gunz *et al.* (2005), Gunz & Mitteroecker (2013) and Botton-Divet *et al.* (2016). 12 juvenile specimens were excluded from the GMMs analyses due to their missing epiphyses (Suppl. Table 1). In order to obtain a comparable set of data, we used the same landmark setup as described in Bader *et al.* (2024): 14 anatomical landmarks for the humerus, 10 for the radius, 18 for the ulna, 17 for the femur, 17 for the tibia and 12 for the fibula (Bader *et al.* 2024, figs S1–S6; tables S2–S7). Each curve is bordered by anatomical landmarks as recommended by Gunz & Mitteroecker (2013). All landmarks and curves were placed using the IDAV Landmark software (v. 3.0, Wiley *et al.* 2005). Following the methodology described in Bader *et al.* (2023) and Bader *et al.* (2024), surface semi-landmarks were placed manually on a template for each type of bone; these templates were then used to project the semi-landmarks onto the surface of the other specimens of the dataset.

## Generalized Procrustes analyses

All specimens were superimposed using a Generalized Procrustes Analysis (GPA) (Rohlf & Slice 1990; Bookstein 1992) to remove the effects of position, orientation and size and to isolate the shape information (3D landmarks coordinates). Specimens' distribution in the morphospace were visualized using Principal Components Analyses (PCA) for all bones except the radius for which no adult dwarf specimen was available. Landmark digitization error was assessed by repeatability tests: ten recordings of anatomical landmarks were made on three visually similar specimens of the same species and analysed by principal component analysis. In order to maximize potential intra-operator error, the landmarks were placed in two sessions of five measurements separated by several days. All repeated measurements produced well-separated clusters on the first two Principal Components (PCs), indicating that measurement error was negligible compared to the biological differentiation among the specimens. Patterns of shape variation were visualized using PCAs computed on each type of bone. PCAs were performed on all specimens, i.e. dwarf, non-dwarf, adult and non-adult. To visualize shape deformation along the principal axes, we computed theoretical consensus shapes from our sample and used them to calculate TPS deformations of the template meshes. The consensus mesh was subsequently used to compute theoretical shapes associated with the maximum and minimum of both axes of each PCA, as well as mean shapes of each bone for each genus. We checked for shape variation difference between genera using Procrustes ANOVAs (Klingenberg 2016) and pairwise comparisons tests (Collyer *et al.* 2015) on the entire shape data provided by the GPAs. GMMs procedures were performed with the 'geomorph' (v. 3.0.7, Adams & Otárola-Castillo 2013; Schlager 2017) and 'Morpho' (v. 2.6) packages of R (v. 4.0.2, R Core Team).

## Allometry

In order to investigate the link between size and shape in the six long bones, we checked for allometry among our sample. Allometry can be defined as the covariation of size with shape (Gould 1966; Klingenberg 2016). To investigate the possible presence of paedomorphic patterns in dwarf elephants, we first examined the growth patterns in extant elephants by testing the ontogenetic allometry (covariation of size with shape during growth) among the *E. maximus* and *L. africana* samples, respectively, with Procrustes analyses of variance (Procrustes ANOVAs; allowing the use of morphometric shape data) using the procD.lm function in the 'geomorph' library (Klingenberg 2016); we then qualitatively compared the morphology of dwarf elephants to the juvenile and adult specimens of extant elephants. In order to determine how dwarf elephants scale compared to non-dwarf species, we checked for evolutionary allometry (covariation of size with shape among different clades at similar ontogenetic stages) among the entire adult sample (dwarf and non-dwarf species) and computed the allometric slopes, allowing for visualizations of the allometric relationship between dwarf and non-dwarf elephants. Finally, we estimated the effect of the size and robustness parameters within the PCAs using linear regressions on the first two PCs, in order to quantify the impact of allometry in the morphological diversity of the sample.

## Robustness analyses

To assess the robustness of the long bones, we measured the maximal length and minimum diaphyseal circumference of each bone. Our sample includes a large number of juvenile specimens, thus often lacking epiphyses. Following Bader *et al.* (2024), we chose to measure the maximal length of the diaphysis as a proxy of the entire bone maximal length (*MaxL*). *MaxL* was defined on each of the six bones as the distance between the proximal and the distal limits of the diaphysis as defined by specific homologous landmarks. Bone maximal diaphyseal length (*MaxL*) and minimal diaphyseal circumference (*Ci*) were obtained virtually by using CloudCompare (version 2.12.0, http://www.cloudcompare.org); robustness (*Rb*) was defined as the ratio of minimal diaphyseal circumference to maximal length of the bones (*Ci/MaxL*). *MaxL* varies greatly between juveniles, adults, dwarf and non-dwarf specimens (e.g. femur; juvenile dwarf: 5.6-7.4 cm; adult dwarf: 35.1 cm; juvenile non-dwarf: 17-58.2 cm, adult non-dwarf: 55.8-73.6 cm). The difference, accounting for ontogenetic stage, in centroid size, bone length, circumference and robustness between genera was tested with ANOVAs. While the small number of dwarf specimens did not allow for quantitative comparison of the robustness between dwarf and non-dwarf specimens, they were however included in the boxplot visualizations of the robustness values, allowing for qualitative comparisons of the robustness variation among the entire sample.

# RESULTS

## Microanatomy

In the six long bones of the dwarf elephants examined, cortical thickness is consistently greater around the growth center (GC), and tapers towards the extremities, typically resulting in a thin layer of compact bone. This distribution forms an hourglass shape in the humerus, ulna and tibia, and to a lesser extent in the femur. In the radius and the fibula, the hourglass shape is very slight. Regardless of the presence of a medullary cavity, trabeculae are always thin and densely packed in the epiphyses and become sparser and thicker around the growth center.

In dwarf elephant calves, cortical thickness is relatively greater compared to the total bone length and diaphyseal width compared to adults for a same given length. The trabeculae in calves are also more isotropic, exhibiting orthogonal anisotropy only in the extremities.

### Humerus

In adult P. tiliensis, the humeral GC is in the distal diaphysis (Fig. 2A, B). The cortex is thicker medially and caudally throughout the diaphysis, and locally thicker in the proximal epiphysis at the greater trochanter and the proximal and medial parts of the head (Fig. 2H). The medullary area is almost entirely filled with trabeculae, with only a few centimeters free above the GC (Fig. 2A, E). Epiphyses are entirely filled with thin trabeculae forming denser zones at articular contacts, such as the humeral head and trochlea (Fig. 2G, H).

In the diaphysis, trabeculae form anisotropic arches oriented from outer regions to the center (Fig. 2A). These arches are oriented proximally above the GC and distally below it, with longer proximal arches reflecting hourglass asymmetry. Above the arches, trabeculae are vertically oriented in the proximal diaphysis (Fig. 2A, B). In the proximal epiphysis, trabeculae are highly anisotropic, oriented distomedially to proximolaterally in the greater trochanter and vertically in the medial and cranial head and lateral metaphysis (Fig. 2H). In the distal epiphysis, they are vertically oriented in the condyles and epicondyles, orthogonal to articular surfaces with the ulna and radius (Fig. 2G).

In the P. falconeri neonate/fetus, the GC is located mid-diaphysis, with a thicker medial cortex, especially distally (Fig. 2C, D). The medullary cavity is fully filled with thick trabeculae forming arches in the diaphysis, though less marked due to lower relative anisotropy (Fig. 2C, D, F). As in adult P. tiliensis, trabeculae are slightly vertically oriented in the outer diaphysis. In the metaphyses, they are highly anisotropic at the extremities of the epiphyses, oriented orthogonally to future epiphyseal lines (Fig. 2C, D).

## Radius

In adult *P. tiliensis*, the cortex is relatively thick (Fig. 3A, B). Cortical thickness is relatively thick (Fig. 3A, B) and homogeneous mediolaterally and craniocaudally, with thinner compact bone at the epiphyses (Fig. 3G, H). The medullary area is filled with dense, highly anisotropic trabeculae (Fig. 3E). This vertical orientation is similar in both epiphyses (Fig. 3G, H), with trabeculae orthogonal to the articular surface with the humerus (Fig. 3G).

The *P. falconeri* neonate/fetus exhibits a similar microanatomical organization (Fig. 3C, D, F), although with less marked trabecular anisotropy. In the diaphysis the trabeculae are only slightly oriented vertically; in the epiphyses they are isotropic except for the most proximal and distal parts where they are oriented orthogonally to the future epiphyseal lines (Fig. 3C).

## Ulna

In adult *P. tiliensis*, the ulnar GC is in the distal third of the diaphysis (Fig. 4A, B). The cortex is relatively thin and forms an asymmetrical hourglass, thicker caudally in the proximal half and medially in the distal half. Compact bone is thin in the epiphyses but locally thicker in the trochlea and styloid process (Fig. 4F, G). The medullary area is partially filled with trabeculae, resulting in a medullary cavity from just below the proximal epiphysis to below the mid-diaphysis. In the epiphyses trabecular trabeculae are relatively thick in the inner part and thinner in the outermost parts. They are highly anisotropic in the olecranon (Fig. 4A, B) and under the trochlea where they are orthogonal to the contact area with the humeral trochlea. Below the medullary cavity the medullary area is filled with distally oriented trabecular arches, followed by vertically oriented trabeculae. The trabecular bone in the distal epiphysis is denser at the contact area with the tarsal bones.

In the *P. falconeri* neonate/fetus, the cortex is thicker caudally but not medially (Fig. 4D, E). The medullary area is filled with trabecular bone, although less dense around the GC (Fig. 4I). The trabeculae are anisotropic although much less marked than in the *P. tiliensis* specimen; oriented vertically in the outermost parts of the bone but mostly isotropic in the inner parts.

## Femur

In adult *P. tiliensis*, the femoral GC is above the mid-diaphysis. The cortex is relatively thick (Fig. 5A, B) and remarkably symmetrical mediolaterally and craniocaudally. It is relatively thin in the epiphyses although locally thicker in the greater trochanter and the proximal part of the femoral neck (Fig. 5J, K). The medullary area is mostly free of trabecular bone from below the GC to the distal third of the diaphysis (Fig. 5A, B). Above the GC, the medullary area is filled with thick, sparse trabeculae (Fig. 5G); and distally with arches of thick trabeculae oriented from the outermost part of the diaphysis towards the center. In the proximal epiphysis, the thin trabeculae are much denser in the femoral head and the medial part of the neck than in the greater trochanter (Fig. 5J, K). They are mostly isotropic in the femoral head, except in the outermost part where the trabeculae are orthogonal to the bone surface. They are oriented vertically in the medial side of the femoral neck, but only slightly anisotropic in the lateral part of the proximal epiphysis (Fig. 4J). In the distal epiphysis, the trabeculae are oriented vertically and are thinner and denser near the compact bone. The anisotropy is more marked medially than laterally; overall, the trabeculae in the trochlea are isotropic (Fig. 5K).

In the calf and neonate/fetus of *P. falconeri*, the GC is in the middle of the diaphysis (Fig. 5C, D, E, F). NMB-Ty.12557 (fetus/neonate) is younger than NMB-Ty.12952 (calf); its cortex and trabeculae are relatively thicker. The hourglass shape of the cortex is more pronounced in both calves than in the adult *P. tiliensis* specimen. In NMB-Ty.12557, the cortex is thicker laterally, while in NMB-Ty.12952, it is more symmetrical and much thicker at the lesser trochanter (Fig. 5C, E, F). Unlike the adult *P. tiliensis* specimen, in the *P. falconeri* juveniles the medullary area is filled with trabecular bone except at the GC (Fig. 5H, I). In the diaphysis the trabeculae form arches similar to those observed in *P. tiliensis*. The proximal and distal extremities show mostly isotropic bone, except at the most proximal and distal areas where the trabeculae are oriented vertically (Fig. 5C, D).

## Tibia

In adult *P. tiliensis*, the tibial GC is in the middle of the diaphysis. The relatively thin cortex surrounds a wide medullary area in NHMW-Geo-1976/1833/0003; in NHMW-Geo-1976/1833/0002 the cortex is relatively much thicker with a more marked hourglass-shaped distribution resulting in a relatively slimmer medullary area (Fig. 6A, B, C, D) but not a larger diaphysis. The cortex is symmetrical mediolaterally but much thicker cranially (Fig. 6). Compact bone is thin in the epiphyses although with local thickenings in the intercondylar crest and the malleolus (Fig. 6H, I).

The medullary area is filled in NHMW-Geo-1976/1833/0002 and almost entirely filled with trabeculae, except at the GC, in NHMW-Geo-1976/1833/0003 (Fig. 6E, F). The trabeculae are thicker and less dense in NHMW-Geo-1976/1833/0003 than in NHMW-Geo-1976/1833/0002. In both specimens they are isotropic in the inner part and highly anisotropic in the outermost parts of the proximal epiphysis, especially below the condyles (Fig. 6H). Under the cranial crest, the trabeculae are oriented caudally and distally (Fig. 6G). Trabeculae are highly anisotropic, oriented vertically in the diaphysis, and form arches from the outermost to the innermost part of the bone in the proximal and distal parts of the diaphysis (Fig. 6C). In the distal epiphysis, the trabeculae are oriented vertically at the tibial cochlea; their anisotropy is lower at the contact area with the fibula where they are oriented orthogonally to the contact area (Fig. 6I). The anisotropy of the epiphyses is more pronounced in NHMW-Geo-1976/1833/0002 than in NHMW-Geo-1976/1833/0003.

No juvenile was available for microanatomical description.

## Fibula

In the fibula of adult *P. tiliensis*, the location of the growth center (GC) is difficult to ascertain. The cortex is relatively thick, a slightly thicker laterally and cranially (Fig. 7A, C, E, F). The compact bone is thicker in the fibular head than the distal epiphysis (Fig. 7B, D, F, G). The medullary area is mostly filled with trabecular bone, although in the middle of the diaphysis, the trabeculae are so thick and sparse that it almost resembles a slim medullary cavity (Fig. 7H, I). The epiphyses contain thinner and denser trabeculae especially in the proximal extremity of the head and at the contact area for the astragalus (Fig. 7J, K). The trabecular bone is slightly anisotropic and oriented vertically in both the diaphysis and the epiphyses, especially in the distal one (Fig. 7A, B). The trabeculae are oriented orthogonally to the contact area with the astragalus in both specimens (Fig. 7J, K). No juvenile was available for microanatomical description.

# Bone cartographies: patterns of cortical thickness distribution

## Adult pattern

To investigate the cortical thickness distribution pattern in adult elephantids, we compared adult specimens of the dwarf species *Palaeoloxodon tiliensis* to those of an extant (non-dwarf) elephant, *Loxodonta africana.*

The humerus of *P. tiliensis* and *L. africana* shows a similar cortical thickness pattern, though *P. tiliensis* has relatively larger areas of high thickness (Fig. 8). In both, compact bone is notably thicker under the deltoid crest (*deltoid* muscle attachment, Shindo & Mori 1956) and distal to the humeral crest (*brachiocephalus* muscle attachment, Trenkwalder 2013). The *P. tiliensis* radius has a relatively thicker cortex than in *L. africana* (relative to diaphyseal width). In both, the cortex is thickest on the medial and lateral diaphysis, but thinner at the contact zone with the ulna (Fig. 8). Conversely, the ulna has thicker compact bone along the radius-contact zone, more pronounced in *L. africana* but extending farther distally in *P. tiliensis*. Both species show thicker bone at the caudal crest for digital flexors attachment (Shindo & Mori 1956). The femur shows the least variation in cortical thickness along its length. It is also the only bone where variation is more marked in *L. africana* than in *P. tiliensis* (Fig. 8). In *P. tiliensis*, the cortex is thickest proximal to the lateral epicondyle (*biceps femoris* attachment). In *L. africana*, though overall variation is slight, the contrast between diaphyseal and epiphyseal thickness is greater, with thicker bone under the femoral neck and greater trochanter. Both show the thickest cortex at mid-diaphysis, especially medially, corresponding to the attachments of the *vastus medialis* and *adductor magnus* muscles. In the tibia, *L. africana* has a thicker cortex at the fibula-contact area, unlike *P. tiliensis*. However, *P. tiliensis* shows thickening at the medial crest (*flexor digitorum communis longus* attachment), while *L. africana* has thicker bone at the cranial crest (*quadriceps femoris* attachment). The fibula shows the most interspecies variation. In *P. tiliensis*, cortical thickness variation between diaphysis and epiphyses is more pronounced than in *L. africana*, with particularly thick cortex on the cranial fibular neck and distocaudal diaphysis, which are areas not associated with muscle attachments.

### Juvenile pattern

To investigate the cortical thickness distribution pattern in juvenile elephantids, we compared juvenile specimens of the dwarf species *Palaeoloxodon falconeri* to those of an extant (non-dwarf) elephant, *Loxodonta africana*. The humerus of both species shows a relatively thicker cortex at mid-diaphysis than in the rest of the shaft; this cortical thickening is much more homogeneous circumferentially in *P. falconeri* than in *L. africana* in which it is more intense on the mediocranial side (Fig. 9). The radial cortex of in *L. africana is* relatively thicker in the proximal half of the shaft and is the thickest at mid-diaphysis. In *P. falconeri* the cortex is relatively thicker and distributed more homogeneously along the shaft. In both species, the compact bone is the thickest on the craniomedial side (Fig. 9). In *P. falconeri*'s ulna, cortical thickness is relatively homogenous along the shaft, with local thickenings under the coronoid processes and the middle of the caudal crest. *L. africana* differs, with a relatively thin cortex and two distinct thickenings under the coronoid processes, more pronounced laterally (Fig. 9). The femur has a thicker cortex at mid-diaphysis in both species. In *P. falconeri*, this thickness is more evenly distributed circumferentially than in *L. africana* in which the thickening is limited to the craniomedial side. In *P. falconeri* compact bone is the thickest at the lesser trochanter; this area is relatively thin in in *L. africana* (Fig. 9). No tibia or fibula of *P. falconeri* were available for comparison.

## Allometry

Results of the Procrustes ANOVAs on the shape data with the centroid size as an independent variable indicate a significant allometry among the entire sample for each bone except the tibia (Table 1). Removing the dwarf specimens, the allometry is significant for each bone, indicating that the allometry present in our sample is not linked to the presence of dwarf specimens only.

## Ontogenetic allometry

Procrustes ANOVAs on the shape data of the non-dwarf sample (adult and juvenile specimens) indicated a significant allometry for each bone. Extant elephants being the only species with several juvenile specimens, all analyses of the shape variation during ontogeny were performed on *E. maximus* and *L. africana* specimens, respectively. The sample size is too small for reliable quantitative analyses of the ontogenetic allometry. Nevertheless, we performed the tests for information only. Despite the limited sample size, we find that in *E. maximus*, shape significantly varies with ontogeny in the humerus, radius, ulna, and femur, whereas it only does so in the humerus in *L. africana* (Suppl. Table 2).

The humerus of juvenile specimens of both *E. maximus* and *L. africana* are very close morphologically, but show distinct growth patterns. In both species, the morphological variation of the humerus during ontogeny is characterized by the development of the epiphyses from small, poorly defined structures into large, well-defined ones (Fig. 10), and by a robustness decrease. However, in *E. maximus*, the humerus develops into a much more robust bone compared to the relatively slimmer humerus of *L. africana*. The decrease in robustness with growth is thus much more pronounced in *L. africana* than in *E. maximus*.

Similarly for the humerus and femur (see Bader *et al.* 2023), the growth of the radius and ulna is characterized by the development of both epiphyses, which is more pronounced in the ulna due to the growth of the massive olecranon (Fig. 11). Additionally, the anconeal process has not yet developed, suggesting a more open elbow, in calves. Both the radius and ulna become less robust with growth, showing relatively slimmer epiphyses; the radius is less curved, and the ulna is less laterally curved.

## Evolutionary allometry

Testing for allometry among the whole sample may allow to visualize how dwarf specimens compare to both adults and juveniles of other elephantid taxa by comparing their allometric slopes. However, the limited number of dwarf specimens does not allow for definitive comparisons (Suppl. Fig. 1). Another way to investigate potential paedomorphic patterns in the shape of limb long bones of dwarf elephant is to compare them qualitatively with those of a non-dwarf of the same genus (here *P. antiquus*) and with juveniles of a close species (here *L. africana*).

We first explored the Interspecific shape variation among Palaeoloxodon by comparing *P. tiliensis* and *P. antiquus*.

In the humerus, the cranial part of the greater trochanter is cranially oriented in *P. tiliensis* but medially oriented in *P. antiquus*, resulting in a relatively larger and more open intertubercular groove in *P. tiliensis* (Fig. 12A). Despite their similar morphology, *P. tiliensis* has a relatively larger humeral head and trochlea. Although the absence of the distal epiphysis on the radius of *P. tiliensis* prevents conclusive comparison, the proximal epiphysis, or radial head, is relatively much larger in *P. tiliensis* than in *P. antiquus* (Fig. 12B). Additionally, *P. tiliensis* shows a relatively wider diaphysis along the craniocaudal axis. The ulna of *P. tiliensis* appears less robust than that of *P. antiquus*, due to its slimmer diaphysis and despite its relatively large epiphyses (Fig. 12C). In *P. tiliensis*, the olecranon is medially oriented, where it is oriented proximomedially in *P. antiquus*. Despite similar trochlear notch depths, the anconeal process is more developed cranially in *P. tiliensis*. The femur is less robust in *P. tiliensis* compared to *P. antiquus*, with a relatively thinner femoral neck supporting a more medial head (Fig. 12D). Additionally, *P. tiliensis* shows a much-reduced anteversion of the femoral neck and relatively less developed greater and lesser trochanters, while the condyles are only slightly relatively smaller than those of *P. antiquus*. The tibial morphology is very similar between the two species, despite the clear microanatomical differences. In *P. tiliensis*, the epiphyses are slightly wider, with a larger facet for the head of the fibula, and the intercondylar eminence extends farther proximally in *P. antiquus* (Fig. 12E). The distal epiphysis in *P. tiliensis* has a steeper angle in the proximodistal axis. The tibial cochlea is deeper in *P. tiliensis*, with a more prominent distal border of the epiphysis. Consistently with the relatively larger facets on the tibia, the proximal and distal contact areas of the fibula with the tibia are relatively larger in *P. tiliensis* than in *P. antiquus* (Fig. 12F). Similarly, the facet angle for the tarsal bones is steeper in *P. tiliensis*, forming a deeper groove.

We then explored the interspecific variation among juveniles by comparing *P. falconeri* and *L. africana.*

The humerus of the juvenile *L. africana* is markedly less robust than that of *P. tiliensis*, with slimmer epiphyses. The deltoid crest is located more distally in *L. africana*, at mid-diaphysis (Fig. 12A). *P. tiliensis* shows a relatively larger humeral trochlea in all axes. Despite the absence of a distal epiphysis in the radius of *P. tiliensis*, the two species shows distinct morphologies: the radial head is relatively much larger in *P. tiliensis* than in *the juvenile L. africana* (Fig. 12B), with a deeper proximal surface. Additionally, *P. tiliensis* displays a relatively wider diaphysis along the craniocaudal axis. In the ulna, the olecranon tuberosity reaches farther medially in the juvenile *L. africana*; in addition, its trochlea is relatively deeper and larger in all directions (Fig. 12C). The femur of the juvenile *L. africana* displays a relatively much thicker neck, associated with a relatively larger greater trochanter reaching more proximally (Fig. 12D). The fossa trochanterica is deeper than in *P. tiliensis*. The diaphysis is more curved medially in *L. africana*; and in the distal epiphysis the condyles are relatively smaller and oriented more distally. The tibia of *P. tiliensis* shows a more robust morphology (large diaphysis and wide epiphyses). The cranial crest is relatively more developed than in *L. africana* (Fig. 12E). Both area of contact with the fibula are relatively much larger in *P. tiliensis* (Fig. 12E). The fibular epiphyses are correspondingly much larger in *P. tiliensis*; the diaphysis is more curved and relatively larger. Finally, the angle of the articular surface for the carpal bones is less steep in *P. tiliensis* than in the juvenile *L. africana* (Fig. 12F).

## Morphological diversity

### Humerus

The first two axes of the PCA performed on humeral shape data of the whole sample express 32.4% of the global variance (Fig. 13A).

The first axis (PC1=17.4%) separates, with some overlap, the taxa according to ontogenetic stage ($p<0.01$, $r^2=0.29$, see Suppl. Table 3); PC1 is significantly correlated with size (*Cs*: $p<0.01$, $r^2=0.48$; *Ci*: $p<0.01$, $r^2=0.47$; *MaxL*: $p<0.01$, $r^2=0.43$) and with robustness ($p<0.01$, $r^2=0.21$). The juvenile specimens are in the positive part of the graph while the adult and subadult specimens are distributed in the center and the negative part. The theoretical shape at PC1 maximum shows a juvenile morphology, with greatly reduced epiphyseal structures (e.g. greater trochanter, humeral trochlea) but relatively large epiphyses compared to the shaft. The PC1 minimum shows instead a large greater trochanter and a clearly defined trochlea, both structures greatly developed in the proximodistal axis.

The second axis (PC2=15.0%) appears to be linked to humeral robustness, although the correlation between PC2 and robustness is not significant ($p=0.05$, $r^2=0.07$, see Suppl. Table 4); PC2 minimum corresponds to an elongated shaft with narrow epiphyses; the epicondyles appear flattened in the mediolateral axis. At the opposite, PC2 maximum corresponds to robust specimens, with a relatively larger diaphysis and even wider epiphyses. The lateral epicondyle is relatively much shorter than at PC2 minimum, and the humeral trochlea is asymmetrical, larger medially. All juvenile specimens are in the positive part of the graph.

## Ulna

The first two axes of the PCA performed on ulnar shape data express 34.8% of the global variance (Fig. 14A). The first axis (PC1=18.2%) is linked to robustness ($p=0.01$, $r^2=0.24$, see Suppl. Table 3) and globally separates *Elephas* specimens from *Mammuthus* and *Palaeoloxodon* specimens, while *Loxodonta* specimens are in the middle of the graph, largely overlapping with the other three genera. PC1 also separates the juvenile specimens of *Elephas* and *Loxodonta* in the negative part (correlation with the ontogenetic stage: $p=0.02$, $r^2=0.15$), and the sole dwarf specimen at the extreme positive part of the graph.

The second axis (PC2=16.6%) separates the juvenile specimens and the dwarf specimen from most of the sample, with the exception of some adult *Elephas* specimens, also in the negative part of the graph. The theoretical shape at PC2 appears to be linked with the orientation of the lateral coronoid process, which extends more distally in juveniles and dwarves than in adult, non-dwarf specimens. However, this coronoid process was partially broken in the dwarf specimen; comparisons with dwarf *P. tiliensis* pictured in the literature (Busk 1868, Scarborough 2020) indicate that this process is not angled down in unbroken ulnae, and suggest that the dwarf specimen, if not damaged, should probably occur in the positive part of PC2. PC2 is significantly correlated with size (*Cs*: $p<0.01$, $r^2=0.23$; *Ci*: $p<0.01$, $r^2=0.25$; *MaxL*: $p<0.01$, $r^2=0.32$) and the ontogenetic stage ($p<0.01$, $r^2=0.22$).

## Femur

The first two axes of the PCA performed on femoral shape data express 37.3% of the global variance (Fig. 15A). The first axis (PC1=22.1%) is linked with robustness ($p<0.01$, $r^2=0.27$, see Suppl. Table 3). The four genera, including adult and juvenile specimens, are widely distributed along PC1; the non-dwarf *Palaeoloxodon* specimens are in the negative part while the dwarf specimen is among the adults of the other genera.

The second axis (PC2=15.2%) separates most juvenile specimens from the rest of the sample, in the positive part of the graph; this axis is significantly correlated with size (*Cs*: $p=0.03$, $r^2=0.10$, *Ci*: $p=0.04$, $r^2=0.09$; *MaxL*: $p=0.04$, $r^2=0.09$) and with the ontogenetic stage ($p<0.01$, $r^2=0.43$). Accordingly, the theoretical shape at PC2 maximum corresponds to a juvenile morphology: a thick femoral neck and an ill-defined trochlea associated with relatively small condyles. The dwarf specimen is in the negative part of the second axis, which correspond to adult specimens.

## Tibia

The first two axes of the PCA performed on tibial shape data express 51.6% of the global variance (Fig. 16A). The first axis (PC1=37.6%) is linked to robustness ($p<0.01$, $r^2=0.37$, see Suppl. Table 3), and separates *Loxodonta* specimens in the negative part from the *Palaeoloxodon* specimens in the positive part of the graph; *Elephas* and *Mammuthus* specimens are in the center, with *Mammuthus* displaying a wider distribution. Juvenile non-dwarf specimens largely overlap with adult ones.

The second axis (PC2=14.0%) appears to be linked to the relative size of the epiphyses but not of the diaphysis: the theoretical shape at PC2 minimum displays large epiphyses compared to PC2 maximum. All juvenile specimens are in the positive part of the second axis, although largely overlapping with adult specimens. The dwarf *Palaeoloxodon* specimen is among adult *Mammuthus* and *Palaeoloxodon* specimens in the positive part of PC1 and the negative part of PC2, which corresponds to a stout morphology.

## Fibula

The first two axes of the PCA performed on fibular shape data express 43.5% of the global variance (Fig. 17A). The first axis (PC1=29.6%) correlates with size (*Cs*: $p=0.04$, $r^2=0.16$; *Ci*: $p<0.01$, $r^2=0.32$, see Suppl. Table 3) and appears to be linked with the relative size of the epiphyses: the negative part of the graph corresponds to a relatively large fibular head and proximodistally enlarged surfaces of contact for the tarsus bones, and the positive part to relatively smaller epiphyses. Adult non-dwarf specimens are widely distributed along the first axis, with the juvenile specimens overlapping in the central and positive parts. The two dwarf specimens are in the center of PC1, indicating an in-between morphology.

The second axis (PC2=13.9%) is linked to robustness ($p<0.01$, $r^2=0.35$), adult non-dwarf specimens are also widely distributed on it, with juvenile specimens in the positive part. Both dwarf specimens are in the positive part of PC2, alongside a juvenile *Loxodonta* specimen and a non-dwarf *Palaeoloxodon*.

### Correlation with size and robustness variables

Centroid size (*Cs*) is correlated with diaphyseal circumference (*Ci*) and maximal length (*MaxL*) for each bone: an increase in centroid size is associated with a larger diaphysis and a greater total length (Table 2). An increase in centroid size is also associated with a higher robustness in the humerus, ulna and femur, but not in the radius, tibia and fibula.

## Robustness analyses

Since most juvenile specimens were lacking epiphyses, robustness was calculated using diaphyseal length (see Bader *et al*. 2024, fig. S1). The boxplots on bone robustness depending on genus and ontogenetic stage indicate distinct patterns, although few of them are significant (Fig. 18).

In extant elephants, bones tend to become more gracile during ontogeny, although adult specimens show different robustness trends depending on species and genera (Bader *et al.* 2023, 2024). In adult non-dwarf specimens, robustness increases along the following order: *Loxodonta* < *Elephas* < *Mammuthus* < *Palaeoloxodon*, with the exception of the radius and femur for which *Loxodonta*'s are more robust than Elephas'. This exception might be linked to the absence of epiphyses in the robustness calculations, as the length of the epiphyses may vary between species and genera. Juvenile specimens are generally less robust than adults, except for the radius in *Elephas* and the fibula in *Loxodonta* among extant elephants. Interestingly, the radius of juvenile *Palaeoloxodon* and the tibia of juvenile *Mammuthus* are more robust than those of adults. For juveniles, the robustness distribution of the humerus, ulna, femur, and tibia typically mirrors that of the adult distribution, except for the radius and fibula (juvenile *Elephas* are more robust than juvenile *Loxodonta*, while adult *Elephas* are less robust than adult *Loxodonta*).

In the dwarf species of *Palaeoloxodon*, both adult and juvenile humeri are less robust than those of non-dwarves. The ulna and femur of juvenile dwarves are more robust than those of juvenile non-dwarves; however the ulna and femur of adult dwarves are less robust than those of non-dwarves. The tibia in dwarf specimens shows a robustness similar to non-dwarf adults, only slightly less robust; the fibula appears slightly more robust in dwarves *Palaeoloxodon* than in non-dwarf *Palaeoloxodon*. Overall, dwarf limb bones tend to be less robust, with the exception of the fibula and no information regarding the radius.

# DISCUSSION

## Dwarf elephants, or the release of weight-bearing constraints

Graviportal proboscideans exhibit numerous anatomical adaptions in their limb bones linked to heavy weight-bearing. These adaptations concern the external morphology, with robust bones with wide epiphyses, ensuring large areas of contact between the bones to maximize the transfer of the mechanical load as well as to maximize the joint stabilization, but also the microanatomy of the bones, with a relatively thick cortex and a medullary area almost entirely filled with advantageously aligned trabecular bone. Our results indicate that while dwarf elephants have retained an overall distinctly proboscidean morphology (see Fig. 19), their limb long bones show clear differences with those of non-dwarf elephant species, both in their shape and their microanatomy, reflecting the reduction of the weight bearing constraints.

## Forelimb bones

The humerus of the dwarf species *P. tiliensis* exhibits a morphology very similar to that of non-dwarf elephant species. However, certain features differ: in *P. tiliensis*, the cranial part of the greater trochanter is oriented cranially, whereas in *P. antiquus*, it is oriented medially. Several muscles attach in that area: the supraspinatus (extensor and stabilizer of the shoulder), the teres minor (adductor, rotator of the forelimb and shoulder stabilizer) and the pectoralis profundus (thorax supporter) (Shindo & Mori 1956). A more cranial orientation of the greater trochanter, combined with the more medial orientation of the humeral head, suggest that in the dwarf species, the humerus is positioned more laterally relative to the body, resulting in a more medial weight distribution. While the humeral cortex in *P. tiliensis* is thicker medially than laterally, supporting the idea of a more medial weight distribution, this pattern also occurs in extant elephants (Bader *et al.* 2025) and is thus not specific to dwarf elephants. If the humerus is indeed placed more laterally in *P. tiliensis*, the more medial load distribution is in fact not particularly reflected in the microanatomy. The absence of clear microanatomical features linked to the variation in load distribution between dwarf and non-dwarf elephants might be linked with the much lighter body of the dwarf specimens. As for trabecular bone, dwarf elephants exhibit relatively highly anisotropic trabeculae in their humerus, which are mostly oriented vertically; while their orientation is similar to those observed in extant elephants, in both the diaphysis and the epiphyses, the anisotropy is overall less pronounced in dwarf elephants. This variation might reflect the reduced constraints due to their lower mass, as is observed in rhinos (same but less marked anisotropy in *Dicerorhinus sumatrensis* than in the four other modern species (Etienne 2023). The lighter weight of dwarf elephant is also reflected in the trabecular density:

although the medullary area remains filled with anisotropic trabeculae, the trabecular bone appears less dense, with larger trabeculae and intertrabecular spaces, indicating a reduced need for weight-bearing adaptations. The humerus of *P. tiliensis* however displays a dense trabecular bone in the middle of the trochlea, corresponding to the zone of contact with the radius. This feature is absent in extant elephants, whose radius is poorly involved in weight bearing.

The radius of *P. tiliensis* shows clear morphological differences compared to non-dwarf species: the radial head and neck are much larger, indicating a greater zone of contact with the humerus and thus a higher involvement in the elbow joint, whether for stabilization or weight support. Consistently, the microanatomy of the radius indicates a much thicker cortex in *P. tiliensis* than in extant elephants (Bader *et al.* 2025). In *Loxodonta africana*, the cortex is thick at mid-diaphysis, forming an hourglass-shaped cone in the bone, whereas cortical thickness is more homogeneous all along the shaft in *P. tiliensis*. This hourglass distribution of the cortical thickness is linked to heavy weight support (Houssaye *et al.* 2021; Etienne 2023), so that its absence suggests a lesser involvement in weight support, whether because the bone is not a primary weight bearer (as is the fibula in extant elephants), or because the weight is not high enough to warrant such a microanatomical adaptation.

The ulna of *P. tiliensis* is distinctly less robust than that of graviportal elephants. This reduced robustness, along with the relatively larger radius, indicates a shift in the relative roles of these two bones. In non-dwarf elephants, and graviportal proboscideans in general, the ulna is the primary weight bearer in the forelimb zeugopod (Smuts & Bezuidenhout 1993; Christiansen 1999). However, while the cortex at the zone of contact with the radius is thicker in the ulna of both extant and dwarf elephants, this thickening extends further distally in *P. tiliensis*, suggesting a higher involvement in load transfer between the two bones and thus a more even distribution of the mechanical load in the zeugopod. Consistently, the medullary area of the ulna is not entirely filled with trabecular bone, suggesting that the overall weight is low enough not to require a fully filled medullary area. In *P. tiliensis* the ulnar olecranon tuberosity is oriented more medially than in other elephantids. This tuberosity bears the attachment of the muscle triceps brachii (a shoulder and elbow extensor); a more medial orientation might suggest a less columnar orientation of the forelimb, with an elbow positioned slightly more laterally than in graviportal species. Additionally, while the trochlear notch is similar in depth in both dwarf and non-dwarf species, the anconeal process is slightly more developed cranially in *P. tiliensis*, indicating a more limited angle of aperture and thus a slightly more bent elbow.

The humeral morphology in *P. tiliensis* suggests a more lateral position of the forelimb than in extant elephants, although the expected increase in mechanical load on the medial side is not reflected in the microanatomy, likely due to the lower body mass of the dwarf species. However, the humerus shows denser trabecular bone at the contact point with the radius, indicating a higher load transfer between these bones compared to in extant elephants. A more laterally placed limb implies a mechanical load and ground reaction forces that are less parasagittally oriented, and thus an increased need for stabilization. The relatively much larger radial head in *P. tiliensis* suggests that its radius is involved in the stabilization of the elbow joint. In addition, the relatively larger size of this bone in *P. tiliensis*, associated with the thicker cortex observed in the ulna where the two bones articulate, suggest a stronger load transfer and thus a more equal distribution of the weight-bearing in the zeugopod bones, as compared to the relatively much larger ulna in non-dwarf graviportal proboscideans. The elbow of *P. tiliensis* thus shows a distinct anatomy from that of extant elephants, reflecting both a lower body mass and the slightly more lateral placement of the forelimb.

## Hindlimb bones

The morphology of the femur of *P. tiliensis* is distinctly different from that of non-dwarf elephant species, especially the proximal epiphysis. In *P. tiliensis*, the femoral head is oriented more medially, indicating that the femur is positioned more laterally to the body, unlike in other elephantids, in which the femoral head is almost directly above the diaphysis, placing the bone directly under the hip joint. Additionally, the femoral neck in *P. tiliensis* is much thinner, reflecting the reduced weight constraints on the femur compared to that of graviportal proboscideans, in which the femoral neck is short and thick, which limits risks of breakage (Bader *et al.* 2024). Similar to the humeral condition, the femur of *P. tiliensis* has a medullary area that is not filled with trabecular bone, suggesting a relaxation of constraints due to body weight. However, unlike in the forelimb, the more lateral position and medial weight distribution are not reflected in the cortical thickness of the femur, which remains remarkably symmetrical along the entire diaphysis, whereas in extant elephants the medial cortex is slightly thicker. Instead, in the femur of *P. tiliensis*, the proximal epiphysis shows a strongly marked anisotropy with vertically oriented trabeculae in the medial part of the head, suggesting that the more medial load bearing is primarily handled by a change in the organization of the trabecular bone. Additionally, the femur of *P. tiliensis* exhibits minimal variation in cortical thickness along the length of its shaft, suggesting that this bone does not require local cortical thickenings to meet its biomechanical constraints. Distally, the caudal orientation of the femoral condyles, instead of the more distal orientation in extant elephants, indicates a more flexed knee and a less columnar posture in dwarf elephants.

The tibia of *P. tiliensis* displays a morphology that closely resembles that of non-dwarf elephant species. However, the intercondylar eminence, which is crucial for knee joint stabilization due to its role in bearing the cruciate ligaments, is less developed cranially in *P. tiliensis*. This suggests a reduced need for joint stabilization compared to in graviportal elephants. Additionally, the articular facets for the fibula are relatively larger in *P. tiliensis*, and the distal articular facet forms a steeper angle, indicating that the fibular distal epiphysis is positioned less directly under the tibia, suggesting a more limited vertical load transfer. These morphological differences suggest that in the dwarf *P. tiliensis* the fibula is more involved as an ankle stabilizer rather than in load transmission.

Compared to in extant elephants, the microanatomy of the tibia of *P. tiliensis* shows a lower anisotropy in the areas of contact with the fibula, reinforcing the hypothesis of a reduced load-bearing role for the fibula. The cortex is the thickest at mid-diaphysis, laterally in *L. africana* and mediocaudally in *P. tiliensis* reflects the more lateral limb position relative to the trunk in the latter. This more lateral limb position and thus more oblique orientation to the mechanical load and ground reaction forces is suggested by the femoral head. These forces being vertical, this might cause these forces to be transferred more medially through the limb. The compact bone is thicker in the tibial cranial crest of *L. africana*, possibly indicating a strong muscle strain from the quadriceps femoris muscle that attaches on that structure; in *P. tiliensis*, despite a thinner compact bone, the high anisotropy in this area suggests that the muscular strain influences its microanatomy. Interestingly, the tibia of *P. tiliensis* does not exhibit a medullary area free of trabecular bone, as might be expected due to its smaller weight, indicating that the filled medullary area was retained during the dwarfism process, unlike in the other limb long bones.
Finally, Scarborough (2020) suggested a more flexible ankle in *P. falconeri* due to the presence of an additional articular facet for the tibia on the calcaneus; however, this facet is not always present in *P. falconeri*, and has not been observed in *P. tiliensis* (C. Tetaert, master's thesis, see Appendix 2), suggesting that this adaptation is not ubiquitous among dwarf elephants. However, Scarborough (2020) also proposed that in *P. falconeri*, the very concave tibial cochlea, extending farther distally on the caudal side, acted as a braking mechanism when walking downslope by limiting displacement on the astragalus. Although less prominent than in *P. falconeri*, the caudal side of the tibial cochlea also extends further distally in *P. tiliensis* than in extant elephants, suggesting a similar role in braking during locomotion on the mountainous terrain of Tilos Island (Nomikou & Papanikolaou 2017).

The fibula of *P. tiliensis* exhibits several distinctive features compared to non-dwarf elephants. It has a relatively larger head and a steeper articular facet for the tibia, as well as a deeper articular facet for the tarsus. The latter aligns with the deeper tibial cochlea, suggesting reduced mediolateral flexibility and enhanced stabilization of the ankle. However, the microanatomy of the fibula presents some inconsistencies with the tibial observations. Unlike in the tibia, in the fibula the area of contact between the two bones is filled with highly anisotropic trabeculae oriented orthogonally to the contact surface, suggesting that the mechanical load on the fibula is strong enough to influence its microanatomy. Indeed, in both *L. africana* and *P. tiliensis*, cortical bone is the thickest on the lateral side, although in *P. tiliensis* it is also thicker at the cranial part of the neck and the caudal part of the diaphysis. These local thickenings strengthen the regions of the maximal curvature along the diaphysis, indicating a greater involvement in weight-bearing than previously expected in the fibula of *P. tiliensis*. However, the absence of corresponding cortical thickening in the tibial facets indicates that this greater relative involvement in load support is not reflected in the microanatomy of the tibia; this discrepancy might be explained by the overall lighter body weight of *P. tiliensis*.

The release of weight-bearing constraints in the dwarf elephant *P. tiliensis* is thus reflected in the limb long bones through a shift towards a more flexed posture, with limbs oriented less vertically than in their mainland ancestor, *P. antiquus*. Additionally, the medullary cavities of *P. tiliensis* contain less trabecular bone, and in both the diaphysis and epiphyses the trabeculae exhibit reduced anisotropy compared to those in extant elephants. Despite the reduction in size, the long bones of *P. tiliensis* still exhibit highly anisotropic, parasagittally vertical trabeculae, suggesting that their body weight necessitate microanatomical adaptations to mechanical loading. Furthermore, the medullary area of *P. tiliensis* remains filled with trabecular bone, highlighting the role of the tibia as the main weight bearer in the hindlimb zeugopod.

## Growth patterns

Organisms that undergo insular dwarfism evolve toward smaller versions of their mainland counterparts. These smaller forms often considered as the result of paedomorphosis; i.e. the retention of juvenile characteristics in adult organisms, which can occur through truncation or elongation of the ontogeny (Gould 1977). It is often considered to be the case in elephants, with dwarf elephant species exhibiting features resembling that of juvenile forms of their mainland ancestors, rather than being isometrically scaled-down versions of the adult forms (Gould 1966, 1977; Accordi & Palombo 1971; Sondaar 1977; Roth *et al.* 1990; Lister 1996). These patterns described in dwarf elephants are mainly linked to the shape of the skull in *P. falconeri*, which exhibit more rounded features, resembling that of juvenile non-dwarf elephants (Palombo 2001), but also to the shape of their humerus and femur (Palombo 2003; Scarborough 2020). However, the presence of paedomorphism among dwarf elephants is debated: recent studies found significant morphological differences between the skulls of dwarf elephants and juvenile mainland elephants, suggesting that the paedomorphic aspect of some features are best explained as a correlate of the smaller skull size (Roth 1984; van der Geer *et al.* 2018). While the histology of dwarf elephant species is scarcely documented, histological analyses of the limb bones and teeth of both *P. falconeri* and *P. cypriotes* indicated that dwarfism was associated with a lower growth rate and an extended growth period (Dirks *et al.* 2012; Köhler *et al.* 2021), without specific mention of paedomorphism in the bones' histology. Bone microanatomy in dwarf elephants have yet to be studied; here, we detail the differences and similarities between dwarf specimens and juveniles of non-dwarf species, before briefly comparing with the general pattern of growth in non-dwarf elephants.

## Do dwarf elephants look like baby elephants?

The bones of the dwarf species *P. tiliensis* exhibit varying degrees of similarity with those of juvenile specimens of extant elephants, revealing a complex pattern of adult and juvenile traits.

The external morphology of the six long bones in adult *P. tiliensis* is very similar to that of adult non-dwarf elephants, with the primary difference being the relative size of the epiphyses. In *P. tiliensis*, the epiphyses are slightly larger than in adult elephants, a trait typical of juvenile non-dwarf specimens. However, aside from this relative size difference, the epiphyses do not exhibit other juvenile characteristics, with the notable exception of the radial head, which is markedly larger in *P.*

*tiliensis* and thus resembles the non-dwarf juvenile morphology. This similarity in relative epiphyseal size between adult *P. tiliensis* and juvenile extant elephant specimens might suggest a paedomorphic trait in the tibia of *P. tiliensis*, which may be linked the increased need for joint stability in a semi-mountainous environment (Sondaar 1977; van der Geer 2014). The microanatomy of the six long bones reveals a similarly mixed pattern. In *P. tiliensis*, the trabecular bone is less dense and has relatively thicker and sparser trabeculae compared to extant elephants, i.e. similar to juvenile elephants' microanatomy. However, although the degree of trabecular anisotropy in the diaphysis and epiphyses of *P. tiliensis* bones is lower than in adult extant elephants, it remains markedly higher than in juveniles of extant elephants. The microanatomy of *P. tiliensis* thus shows some juvenile-like traits, which might be interpreted as resulting from paedomorphism, but retains a level of anisotropy characteristic of adults. This “partial paedomorphic pattern” could better be explained by the size and mass discrepancy between dwarf and non-dwarf elephants: the trabecular bone grows denser with increasing biomechanical constraints (Currey 2002), so that the relatively lesser trabecular density observed in *P. tiliensis* could be linked to its much lighter body compared to extant elephants.

As a result, we find no clear evidence of paedomorphism in the limb long bones of *P. tiliensis.* While our results are consistent with several recent studies (e.g. Roth 1984; van der Geer *et al.* 2018), they are in contradiction with several other studies suggesting paedomorphic morphological patterns in dwarf elephants. These studies, however, were not conducted on *P. tiliensis*: most of them focus on *P. falconeri*, and sometimes include other dwarf elephant species such as *Mammuthus lamarmorai*, *Mammuthus creticus* or *Palaeoloxodon mnaidriensis*. Since the effects of insular dwarfism vary depending on a large array of factors (e.g. island size, presence of competitors, topography) (Sondaar,1977; van der Geer 2014; van der Geer *et al.* 2016, 2018), our observations do not necessarily contradict the results of the previous studies on other species of dwarf elephants, but suggest that *P. tiliensis* shows no indication of paedomorphism in its post-cranial skeleton. In addition, most of these studies were conducted on skull shape data, with the notable exception of Herridge (2010), who investigated the allometric pattern on the limb long bones, and Scarborough (2020) who described the morphology of the whole limb.

The morphological variation between dwarf and non-dwarf elephant species has been linked to the duration of island isolation (van der Geer *et al.* 2016). *P. tiliensis* (1300 kg) has been isolated for approximately 45.000 years (Athanassiou *et al.* 2019), a much shorter period than the several hundred thousand years of isolation of *P. falconeri* (300 kg, Scarborough 2022). Thus, it is unsurprising to find a less derived morphology in *P. tiliensis* than in *P. falconeri*.

## Growth patterns in dwarf elephants

Our sample include fetus/neonate bones (humerus, radius, ulna and femur) of *P. falconeri*, the smallest dwarf elephant that ever existed (Larramendi 2016). Their qualitative comparison (the absence of epiphyses preventing their inclusion in the shape analyses) with the adult *P. falconeri* specimens illustrated by Scarborough (2020) provided valuable insights.

Similar to in extant elephants, the growth of the humerus in *P. falconeri* humerus is characterized by a relatively stronger development of the proximal epiphysis, while the distal epiphysis maintains roughly the same proportions throughout growth. The ulna depicted in Scarborough (2020) shows a massive medial coronoid process, which was not observed in the fetus/neonate examined here (another *P. falconeri* ulna, depicted in Bonfiglio *et al.* (2003) in medial view only, prevents comparison). However, (Busk 1868) made a similar observation, suggesting that this large medial coronoid process is not a pathological anomaly but rather a species-specific characteristic of *P. falconeri,* likely developed later during growth. Finally, we find a clear trend of decreasing robustness with age in *P. falconeri*, i.e. the opposite trend to what is observed in the femur of extant elephants during growth (Bader *et al.* 2023).

Trabeculae are relatively thicker in bones of juvenile *P. falconeri* (300 kg) than in calves of extant elephants (100 kg). This variation might be linked to the absolute size of the bone, i.e. both dwarf and non-dwarf juvenile specimens might have trabeculae of the same absolute size, appearing thinner in non-dwarf specimens due to the largest size of the bone. Further quantitative analyses of the trabeculae would be required to conclude on this point. Unlike in juveniles of non-dwarf elephants, the trabeculae are mostly isotropic in juvenile *P. falconeri* bones, suggesting a lesser need for trabecular anisotropy involved in weight support in the latter. Conversely, this isotropy might instead reflect the youth of the fetus/neonate specimen, which is here compared to a young but older elephant calf.  Although no adult specimens were available for direct comparison, our sample includes another femur from an older juvenile specimen of *P. falconeri*, allowing some insight into the ontogenetic growth pattern in this species. The trabeculae become relatively thinner and more numerous with growth, and the older specimen displays a slight vertical anisotropy, indicating that trabecular anisotropy may increase during growth in *P. falconeri*, despite the species' relatively small body weight. Additionally, the bones of *P. falconeri* exhibit a medullary cavity filled with trabecular bone, similar to that of extant elephant calves. In terrestrial taxa, a filled medullary area is a typical adaptation to support heavy weight (Houssaye *et al.* 2021). The presence of this feature in a fetus/neonate specimen of the smallest dwarf elephant species thus appears unnecessary, and is a remnant of its heavier relatives.

## Insular dwarfism in proboscideans: unique and shared adaptations

### Do dwarf elephants display plesiomorphic traits?

Early proboscideans were relatively small and did not display features associated with graviportal adaptations (Bader *et al.* 2024), which appeared alongside the general size increase in the proboscidean order, ultimately resulting in the graviportal pattern that is now observed in extant elephants. The massive size reduction undergone by dwarf elephants is an excellent case study to investigate if graviportal features are maintained or lost in dwarf species. We found a number of plesiomorphic traits in the limb long bones of dwarf elephants, albeit to varying degrees depending on the bones considered.

The humerus of *P. tiliensis* only shows little similarity with that of early, non-graviportal proboscideans, with the exception of the orientation of the head, which is more medial and placed slightly less above the humeral diaphysis. In addition, the greater trochanter is oriented cranially, so that the intertubercular groove is more open. Together, these features indicate that the limb is placed more laterally to the trunk than in extant elephants, as observed in *Moeritherium* (400 kg) and *Numidotherium* (300 kg). While the humerus of *M. lamarmorai* (550 kg) figured in Palombo *et al.* (2012) resembles that of *P. tiliensis* (1300 kg), the humerus of *P. falconeri* (300 kg) figured in Scarborough (2020) does not show such a medial orientation of the head, suggesting that not all dwarf elephants display this similarity with *Moeritherium*. In addition, in *Moeritherium* as in *P. tiliensis*, the trabeculae of the humerus are oriented vertically in the diaphysis and the epiphyses, although it is less marked than in extant elephants (pers. obs.).

In mainland proboscideans, the radius and ulna are not fused but articulate via a large, flat surface, significantly limiting or preventing pronation and supination movements. This anatomical arrangement may have predisposed dwarf species to evolve complete fusion of the zeugopod bones both proximally and distally. In several dwarf proboscidean species, such as *P. falconeri* (300 kg) and *Mammuthus mnaidriensis* (1700 kg), the radius and ulna are fused (Ambrosetti 1968; Ferretti 2008). However, this fusion is not observed in *P. tiliensis* (Theodorou *et al*. 2007), despite its slightly larger size compared to *P. falconeri*, indicating that this characteristic is not directly, or not only, linked to body mass. The fusion of these bones has been proposed as an adaptation to the more strenuous locomotor activity imposed by the semi-mountainous terrain of Mediterranean islands (Larramendi 2016). Interestingly, this characteristic is absent in the pygmy mammoths of the Channel Islands (Agenbroad *et al.* 1999), but present in one of the dwarf stegodont species of the Indonesian Islands, *Stegodon florensis* (Court 1994; van den Bergh 1999, 2008). Among proboscideans, this trait appears to be specific to dwarf species, indicating a morphological convergence that could be driven by similar environmental pressures; however, it is not present in every dwarf species, indicating varied adaptive strategies to the semi-mountainous island habitats. Interestingly, some early non-graviportal proboscideans, such as *Numidotherium*, also exhibited a fused zeugopod, with the radius and ulna always fused distally and sometimes proximally (Court 1994). The plesiomorphic condition for proboscideans being unknown, fusion observed in both early, relatively small (300 kg) proboscideans and dwarf elephants (ranging from 300 kg to 1500 kg) suggest a morphological convergence, despite their markedly different habitats since *Moeritherium* lived in marshy and fluvial environments (Mahboubi *et al.* 2014, Sanders 2024). In dwarf elephants and in early proboscideans, the fusion of the radius and ulna likely reduced the movement of the proximal carpals relative to the distal ulna-radius, particularly in medio-lateral motion, increasing wrist stability (Court 1994; Scarborough 2020). In addition, as in early, non-graviportal proboscideans, in dwarf elephants the radius appears to be more involved in weight bearing. This appears to also be the case in *P. falconeri*, *P. mnaidriensis* and *M. lamarmorai* (Palombo *et al.* 2012; Scarborough 2020), so that this plesiomorphic trait of a relatively larger radius appears to be present in at least two genera of dwarf elephants, *Palaeoloxodon* and *Mammuthus*.

The femur of *P. tiliensis* shows a smaller anteversion angle, as well as caudally oriented condyles. Together, these features indicate a more laterally oriented and more flexed knee than observed in non-dwarf elephants, and thus a slight reversion of the columnar stance. While still mostly columnar, the posture of the dwarf elephant is reminiscent of the flexed limbs of early, non-columnar proboscideans such as *Numidotherium* or *Moeritherium* (Court 1994; Mahboubi *et al.* 2014). One of the main differences between early proboscideans and graviportal proboscideans is the presence of a visible third trochanter in the former, bearing the attachment of the muscle gluteus superficialis (Barone 2010; Etienne *et al.* 2021); we found no indication that this trochanter was present in *P. tiliensis*, nor does it appear to be present in *P. falconeri* (Scarborough 2020). A slightly more developed third trochanter, compared to that of extant elephants, was however described in dwarf stegodont species from Flores, *Stegodon sondaari* and *Stegodon florensis* (Wibowo 2016), indicating that in those species the less columnar hindlimb compared to continental species was associated with an increased reliance on leg abductor and extensor muscles. *P. tiliensis*' femur also shares with *Moeritherium*'s a medullary area filled with trabecular bone, with trabeculae oriented vertically (pers. obs.) but a generally lower degree of anisotropy than in extant elephants, suggesting a design less dedicated to load bearing. If confirmed, in *Moeritherium* this organization might also be linked its semi-aquatic lifestyle (Liu *et al.* 2008).

The morphology of *P. tiliensis'* tibia closely resembles that of other elephantids', with a notable exception in the relative size of the articular surfaces for the fibula. These relatively larger contact zones were also described in other dwarf elephants, *P. falconeri* and *P. mnaidriensis* (Scarborough 2020). Among early proboscideans, the heavy but non-columnar *Barytherium grave* (2000 kg) is the only taxon with a complete tibia available for comparison, limiting direct comparisons with earlier, smaller taxa, except for an incomplete tibia of *Moeritherium.* The tibia and fibula of *P. tiliensis* articulate via markedly larger contact zones than in elephantids, especially proximally. This condition is much closer to what is observed in both *Moeritherium* and *Barytherium* (pers. obs.), suggesting a reversion of the morphology of the hindlimb zeugopod to a plesiomorphic state, with a relatively much larger fibula, more involved in load bearing.

Through the dwarfism process, dwarf elephants lost some of their graviportal adaptations, exhibiting plesiomorphic traits in their limb bones that resemble those of early, non-graviportal proboscideans. Dwarf elephants however retained key graviportal features compared to these early proboscidean species, including a great overall robustness, generally enlarged articular surfaces, straight diaphyses, and a predominantly columnar posture.

## How do dwarf elephants compare to other heavy dwarves?

Proboscideans are not the only heavy mammals that underwent insular dwarfism; while dwarf rhino species have been scarcely studied, dwarf hippo species have been more largely described, allowing for comparison of dwarfism patterns among heavy animals. Dwarf hippos are currently represented by two species: the large *Hippopotamus amphibius*, generally considered as graviportal (although not displaying a columnar stance), and the smaller pygmy hippo, *Choeropsis liberiensis*, whose relatively small size is plesiomorphic and not due to insular dwarfism (Boisserie *et al.* 2011). The fossil record indicates that some populations of hippos underwent a reduction in size on Cyprus and Madagascar, resulting in the dwarf hippo species *Phanourios minor* (approx. 200 kg), *Hippopotamus madagascariensis* (approx. 400 kg) and *Hippopotamus lemerlei* (approx. 350 kg) (Sondaar 1977; Boisserie *et al.* 2011; van der Geer *et al.* 2018; Palombo & Moncunill-Solé 2023).

Although dwarf hippopotamuses have more robust bones than their living relatives (Prothero & Sereno 1982; Georgitsis *et al.* 2022), comparisons between the three species indicate that dwarf hippos retain more similar morphological features to those of the heavy *H. amphibius* than to those of the pygmy hippo *C. liberiensis*, indicating that dwarfism did not strongly affect the limb bones (Georgitsis *et al.* 2022). Both dwarf elephants and dwarf hippos thus retained a morphology more closely resembling their large relatives than that of their smaller, more distantly related counterparts. Dwarf hippos and dwarf rhinos show significantly more robust bones than their non-dwarf relatives (Prothero & Sereno 1982; Georgitsis *et al.* 2022). This increased robustness is also typically associated with dwarf elephants (e.g. Roth 1990; Ferretti 2008; Herridge 2010), although it is not always the case. The Channel Islands mammoth, *Mammuthus exilis* does not display more robust bones than its mainland ancestor (Htun *et al*. 2018).

Our findings indicate that while *P. tiliensis* indeed have generally more robust bones than non-palaeoloxodont elephants, they exhibit a robustness similar to that of their mainland ancestor *P.*

*antiquus*, and thus do not display an increased robustness associated with their dwarfism. In addition, while dwarf elephants have retained large epiphyses, the dwarf hippo *P. minor* exhibits distinctly slimmer ones, which have been linked to their reduced weight load (Georgitsis *et al.* 2022). This difference in morphological variation during dwarfism underscores that there is no single way for large taxa to adapt to size reduction.

The fusion of the radius and ulna, as well as of the tibia and fibula, observed in *P. minor* is similar to that seen in *P. falconeri* (Ambrosetti 1968; van der Geer *et al.* 2013). In both cases, this fusion has been hypothesized as linked to increased limb stability (van der Geer *et al.* 2013), facilitating locomotion in otherwise inaccessible mountainous terrain (Sondaar 1977). This similarity between phylogenetically distant clades suggests evolutionary convergence during dwarfism, and appears to be driven by locomotor needs in mountainous islands (Athanassiou *et al.* 2019).

# CONCLUSION

This study reveals that while the dwarf elephant *Palaeoloxodon tiliensis* retains key proboscidean features in its limb long bones, it also exhibits numerous anatomical differences compared to its larger relatives. These correspond to variations in both the morphology and the microanatomy of the bones, and reflect the reduced weight bearing constraints in this dwarf species, the latter resulting in a shift toward a more flexed posture and a less parasagittal limb orientation compared to the columnar limbs typically associated with graviportal proboscideans. While the external morphology of the long bones in *P. tiliensis* largely resembles that of adult non-dwarf elephants, their relatively larger epiphyses, especially the radial head, suggest some juvenile-like features. These traits might be adaptive in the context of a reduced mechanical load or of locomotion on a more challenging environment (semi-mountainous) compared to their mainland relatives. Comparisons with early, non-graviportal proboscideans indicate that *P. tiliensis* exhibits several plesiomorphic traits in its limb long bones, to varying degrees depending on the bone considered. However, despite the loss of some graviportal features, *P. tiliensis* retains several key adaptations associated with a graviportal lifestyle, such as the overall great robustness, enlarged articular surfaces, straight diaphyses, and a predominantly columnar posture. These results suggest that, although dwarfism led to a partial reversion to ancestral traits of early, smaller-bodied proboscidean species, *P. tiliensis* still maintains overall graviportal features in its limb bones' morphology. Unlike in extant elephants, the limb long bones of *P. tiliensis* show a medullary area only partially filled with trabecular tissue, further underlining the release of weight-bearing constraints. However, *P. tiliensis* exhibits highly anisotropic, vertically oriented trabeculae in the epiphyses and, to a lesser extent, in the diaphysis, as in extant elephants, which shows that certain features were retained despite dwarfism; similarly, the juvenile *P. falconeri* specimens display a medullary area mostly filled with trabecular bone, as in adult *P. tiliensis* and both juvenile and adult extant elephants. In addition, trabecular anisotropy increases during growth in *P. falconeri*, despite this species small body size, further highlighting weight-support related features maintained despite dwarfism. The bone microanatomy of *P. tiliensis* illustrates this mixed signal, showing a trabecular structure that is less dense and more juvenile-like compared to extant elephants, yet still more anisotropic than in juvenile non-dwarf elephants. This suggests that while dwarf elephants exhibit some juvenile-like characteristics in both their shape and microanatomy, these might also be linked to their lighter body mass and reduced weight-bearing constraints, rather than to a paedomorphic pattern alone. Dwarf elephants are therefore not an isometrically scaled-down version of their mainland ancestors, nor do they totally resemble juveniles of extant elephants nor a complete reversion to early proboscidean traits. Instead, they exhibit unique combinations of graviportal, juvenile, and plesiomorphic traits, reflecting a distinct evolutionary pathway.

# ACKNOWLEDGEMENTS

This work was funded by the European Research Council as part of the GRAVIBONE project (ERC-2016-STG-715300), and has received financial support from the CNRS through the MITI interdisciplinary programs and from the CSTB. The funders had no role in study design, data collection and analysis, decision to publish, or preparation of the manuscript.

We warmly thank J. Lesur, C. Bens, A. Verguin, G. Billet (Muséum national d'Histoire naturelle (MNHN), Paris, France), O. Pauwels, S. Bruaux, J. Brecko (Royal Belgian Institute of natural Sciences, Brussels, Belgium), F. Zachos, A. Bibl (Naturhistorisches Museum Wien (NHMW), Vienna, Austria), L. Costeur (Naturhistorisches Museum Basel, Switzerland), A. van Heteren (Zoologische Staatssammlung München, Munich, Germany), J. Galkin, R. O'Leary, C. Mehling, A. Gishlick and J. Meng (American Museum of Natural History, New-York, USA), N. Gilmore, E. Daeschler, J. P. Downs (Academy of Natural Science of Philadelphia (ANSP), Philadelphia, USA), D. Berthet (Musée des Confluences, Lyon, France), P. Kamminga, S. Vandermije, N. den Ouden (Naturalis Biodiversity Center, Leiden, Netherlands), M. Gasparik, M. Segesdi (Hungarian National History Museum, Budapest, Hungary), R. Tabuce, M. Mouana (Institut des Sciences de l'Évolution, Montpellier, France), L. Makádi (Magyar Állami Földtani Intézet, Budapest, Hungary), E. Cioppi, A. Savorelli, L. Bellucci, S. Dominici, B. Montecchi (Museo di Geologia e Paleontologia, Università degli studi di Firenze, Florence, Italy), N. Adams, R. Pappa, J. R. Hutchinson, L. Tomsett, R. Sabin (Natural History Museum, London, UK), E. Gilissen, A. Mathys (Royal Museum for Central Africa, Tervuren, Belgium), E. Amson, M. Böller (Staatliches Museum für Naturkunde Stuttgart, Germany), R. Narducci, R. Hulbert, A. Woodruff, J. Bourque (University of Florida, Gainesville, USA) for granting access to the specimens. Further thanks to M. Bellato from the AST-RX platform in the MNHN (UMS 2700, Paris, France) and to V. Winkler in the NHMW for performing scans and reconstructions.

Further thanks to M. Amari, P. Costes, N. Mohanty, I. Pelletan, M. Sowinski, K. Stilson and C. Tetaert (MNHN, Paris, France) for their assistance in transporting heavy material, and R. Gilardet for help segmenting some specimens.

We thank the IMNH, the ANSP and the Duke Lemur Center of Natural History (DLCNH) for the raw data and 3D models on MorphoSource: Idaho Museum of Natural History provided access to the IMNH-1486 specimen data, the collection of which was funded by the Rick Carron Foundation.

We also thank two anonymous reviewers for their helpful comments and corrections, as well as Sally Thomas for editorial work.

# AUTHOR CONTRIBUTIONS

**Conceptualization** Camille Bader (CB), Alexandra Houssaye (AH); **Data curation** CB, AH, Ursula Göhlich (UG); **Formal Analysis** CB; **Funding Acquisition** AH; **Investigation** CB, AH; **Methodology** CB, AH; **Project Administration** CB, AH; **Supervision** AH; **Validation** CB, AH, UG; **Visualization** CB; **Writing – Original Draft Preparation** CB**; Writing – Review & Editing** CB, AH, UG.

# FUNDING

This work was funded by the European Research Council as part of the GRAVIBONE project (ERC-2016-STG-715300). The funders had no role in study design, data collection and analysis, decision to publish, or preparation of the manuscript.

# CONFLICT OF INTEREST

The authors declare no conflict of interest.

# DATA ARCHIVING STATEMENT

Raw CT-scan data of the MNHN specimens are archived at the Muséum national d'Histoire naturelle, Paris, France and registered on the 3Dtheque portal: https://3dtheque.mnhn.fr/. Raw CT-scan data of the NHMUK and RVC specimens are registered on figshare: doi:10.6084/m9.figshare.26779570.

# REFERENCES

**Accordi**, F.S., Palombo, M.R., 1971. Morfologia endocranica degli elefanti nani pleistocenici di Spinagallo (Siracusa) e comparazione con l'endocranio di Elephas antiquus. Atti della Accademia Nazionale dei Lincei. Classe di Scienze Fisiche, Matematiche e Naturali. Rendiconti 51, 111–124.

**Adams**, D.C., Otárola-Castillo, E., 2013. geomorph: an R package for the collection and analysis of geometric morphometric shape data. Methods Ecol Evol 4, 393–399. https://doi.org/10.1111/2041-210X.12035

**Agenbroad**, L.D., 2012. Giants and pygmies: mammoths of Santa Rosa Island, California (USA). Quaternary International 255, 2–8.

**Agenbroad**, L.D., Morris, D., Roth, L., 1999. Pygmy mammoths Mammuthus exilis from Channel Islands National Park, California (USA). Deinsea 6, 89–102.

**Aiba**, H., Baba, K., Matsukawa, M., 2010. A new species of Stegodon (Mammalia, Proboscidea) from the Kazusa Group (lower Pleistocene), Hachioji City, Tokyo, Japan and its evolutionary morphodynamics. Palaeontology 53, 471–490.

**Ambrosetti**, P., 1968. The Pleistocene dwarf elephants of Spinagallo (Siracusa, south-eastern Sicily). Geologica romana, 277-398.

**Athanassiou**, A., Van Der Geer, A.A.E., Lyras, G.A., 2019. Pleistocene insular Proboscidea of the Eastern Mediterranean: A review and update. Quaternary Science Reviews 218, 306–321. https://doi.org/10.1016/j.quascirev.2019.06.028

**Bader**, C., Böhmer, C., Abou, M., Houssaye, A., 2022. How does bone microanatomy and musculature covary? An investigation in the forelimb of two species of martens ( *Martes foina* , *Martes martes* ). Journal of Anatomy 241, 145–167. https://doi.org/10.1111/joa.13645

**Bader**, C., Delapré, A., Göhlich, U.B., Houssaye, A., 2024. Diversity of limb long bone morphology among proboscideans: how to be the biggest one in the family. Papers in Palaeontology 10, e1597.

**Bader**, C., Delapré, A., Houssaye, A., 2023. Shape variation in the limb long bones of modern elephants reveals adaptations to body mass and habitat. Journal of Anatomy 242, 806–830. https://doi.org/10.1111/joa.13827

**Bader**, C., Gilardet, R., Rinder, N., Herridge, V., Hutchinson, J. R., & Houssaye, A. 2025. Long-bone microanatomy in elephants: microstructural insights into gigantic beasts. Zoological Journal of the Linnean Society, 204(3), zlaf008.

**Baleka**, S., Varela, L., Tambusso, P.S., Paijmans, J.L.A., Mothé, D., Stafford, T.W., Fariña, R.A., Hofreiter, M., 2022. Revisiting proboscidean phylogeny and evolution through total evidence and palaeogenetic analyses including Notiomastodon ancient DNA. iScience 25, 103559. https://doi.org/10.1016/j.isci.2021.103559

**Barone**, R., 2010. Anatomie comparée des mammifères domestiques. Tome 2, Arthrologie et myologie, 4e édition. ed. ACV, Paris.

**Bertram**, J.E.A., 2016. Design for Prodigious Size without Extreme Body Mass: Dwarf Elephants, Differential Scaling and Implications for Functional Adaptation, in: Bertram, J.E.A. (Ed.), Understanding Mammalian Locomotion. Wiley, pp. 349–367. https://doi.org/10.1002/9781119113713.ch14

**Boisserie**, J.-R., Fisher, R.E., Lihoreau, F., Weston, E.M., 2011. Evolving between land and water: key questions on the emergence and history of the Hippopotamidae (Hippopotamoidea, Cetancodonta, Cetartiodactyla). Biological Reviews 86, 601–625.

**Bonfiglio**, L., Galeani, D., Insacco, G., Mangano, G., Marra, A.C., others, 2003. Elephas falconeri (Busk, 1867) and Leithia melitensis (Adams, 1863) from a karst fissure of the Hyblean Plateau (South-eastern Sicily). Bollettino della Società Paleontologica Italiana 42, 123–128.

**Bookstein**, F.L., 1992. Morphometric Tools for Landmark Data: Geometry and Biology, 1st ed. Cambridge University Press. https://doi.org/10.1017/CBO9780511573064

**Botton**-**Divet**, L., Cornette, R., Fabre, A.-C., Herrel, A., Houssaye, A., 2016. Morphological analysis of long bones in semi-aquatic mustelids and their terrestrial relatives. Integrative and comparative biology 56, 1298–1309.

**Busk**, G., 1868. Description of the Remains of three extinct Species of Elephant, collected by Capt. Spratt, CB, RN, in the Ossiferous Cavern of Zebbug, in the Island of Malta. The Transactions of the Zoological Society of London 6, 227–306.

**Cantalapiedra**, J.L., Sanisidro, Ó., Zhang, H., Alberdi, M.T., Prado, J.L., Blanco, F., Saarinen, J., 2021. The rise and fall of proboscidean ecological diversity. Nat Ecol Evol 5, 1266–1272. https://doi.org/10.1038/s41559-021-01498-w

**Case**, T.J., 1978. A General Explanation for Insular Body Size Trends in Terrestrial Vertebrates. Ecology 59, 1–18. https://doi.org/10.2307/1936628

**Christiansen**, P., 1999. Scaling of mammalian long bones: small and large mammals compared. Journal of Zoology 247, 333–348. https://doi.org/10.1111/j.1469-7998.1999.tb00996.x

**Collyer**, M.L., Sekora, D.J., Adams, D.C., 2015. A method for analysis of phenotypic change for phenotypes described by high-dimensional data. Heredity 115, 357–365. https://doi.org/10.1038/hdy.2014.75

**Coombs**, W.P., 1978. Theoretical Aspects of Cursorial Adaptations in Dinosaurs. The Quarterly Review of Biology 53, 393–418. https://doi.org/10.1086/410790

**Court**, N., 1994. Limb posture and gait in Numidotherium koholense, a primitive proboscidean from the Eocene of Algeria. Zoological Journal of the Linnean Society 111, 297–338. https://doi.org/10.1111/j.1096-3642.1994.tb01487.x

**Currey**, J.D., 2002. Bones: structure and mechanics. Princeton University Press, Princeton, NJ.

**Curtin**, A.J., Macdowell, A.A., Schaible, E.G., Roth, V.L., 2012. Noninvasive histological comparison of bone growth patterns among fossil and extant neonatal elephantids using synchrotron radiation X-ray microtomography. Journal of Vertebrate Paleontology 32, 939–955. https://doi.org/10.1080/02724634.2012.672388

**Dirks**, W., Bromage, T.G., Agenbroad, L.D., 2012. The duration and rate of molar plate formation in Palaeoloxodon cypriotes and Mammuthus columbi from dental histology. Quaternary International 255, 79–85. https://doi.org/10.1016/j.quaint.2011.11.002

**Etienne**, C., 2023. Biomechanical adaptations to support high body weight, an investigation in limb long bones of Rhinocerotoidea. Paris Cité, Paris.

**Etienne**, C., Filippo, A., Cornette, R., Houssaye, A., 2021. Effect of mass and habitat on the shape of limb long bones: A morpho-functional investigation on Bovidae (Mammalia: Cetartiodactyla). Journal of Anatomy 238, 886–904. https://doi.org/10.1111/joa.13359

**Ferretti**, M., 2008. The dwarf elephant Palaeoloxodon mnaidriensis from Puntali Cave, Carini (Sicily; late Middle Pleistocene): Anatomy, systematics and phylogenetic relationships. Quaternary International 182, 90–108.

**Foster**, J.B., 1964. Evolution of Mammals on Islands. Nature 202, 234–235. https://doi.org/10.1038/202234a0

**Genoud**, M., Christe, P., 2011. Thermal energetics and torpor in the common pipistrelle bat, Pipistrellus pipistrellus (Vespertilionidae: Mammalia). Comparative Biochemistry and Physiology Part A: Molecular & Integrative Physiology 160, 252–259. https://doi.org/10.1016/j.cbpa.2011.06.018

**Georgitsis**, M.K., Liakopoulou, D.E., Theodorou, G.E., Tsiolakis, E., 2022. Functional morphology of the hindlimb of fossilized pygmy hippopotamus from Ayia Napa (Cyprus). Journal of Morphology 283, 1048–1079. https://doi.org/10.1002/jmor.21488

**Gould**, S.J., 1977. Ontogeny and phylogeny. Belknap Press of Havard University Press, Cambridge, Mass. London, England.

**Gould**, S.J., 1966. Allometry and size in ontogeny and phylogeny. Biological Reviews 41, 587–638. https://doi.org/10.1111/j.1469-185X.1966.tb01624.x

**Gregory**, W.K., 1912. Notes on the principles of quadrupedal locomotion and on the mechanism of the limbs in hoofed animals. Annals of the New York Academy of Sciences 22, 267–294. https://doi.org/10.1111/j.1749-6632.1912.tb55164.x

**Gunz**, P., Mitteroecker, P., 2013. Semilandmarks: a method for quantifying curves and surfaces. Hystrix, the Italian Journal of Mammalogy 24, 103–109.

**Gunz**, P., Mitteroecker, P., Bookstein, F.L., 2005. Semilandmarks in three dimensions, in: Slice, D. (Ed.), Modern Morphometrics in Physical Anthropology. Kluwer Academic Publishers/Plenum Publishers, New York, pp. 73–98.

**Hayashi**, S., Kubo, M.O., Sánchez-Villagra, M.R., Taruno, H., Izawa, M., Shiroma, T., Nakano, T., Fujita, M., 2023. Variation and process of life history evolution in insular dwarfism as revealed by a natural experiment. Front. Earth Sci. 11, 1095903. https://doi.org/10.3389/feart.2023.1095903

**Herridge**, V.L., 2010. Dwarf elephants on Mediterranean islands: a natural experiment in parallel evolution (PhD Thesis). UCL (University College London).

**Herridge**, V.L., Lister, A.M., 2012. Extreme insular dwarfism evolved in a mammoth. Proc. R. Soc. B. 279, 3193–3200. https://doi.org/10.1098/rspb.2012.0671

**Houssaye**, A., Martin, F., Boisserie, J.-R., Lihoreau, F., 2021. Paleoecological inferences from long bone microanatomical specializations in Hippopotamoidea (Mammalia, Artiodactyla). Journal of Mammalian Evolution 28, 847–870.

**Htun**, T., Prothero, D.R., Hoffman, J.M., Lukowski, S.M., Syverson, V., 2018. Allometric trends in dwarfing in the extinct Pleistocene Channel Islands pigmy mammoth, Mammuthus exilis. Fossil Record 6 Volume 1 79, 261.

**Hutchinson**, J.R., Famini, D., Lair, R., Kram, R., 2003. Are fast-moving elephants really running? Nature 422, 493–494. https://doi.org/10.1038/422493a

**Johnson**, P.M., Delean, S., 2003. Reproduction of Lumholtz's tree-kangaroo, Dendrolagus lumholtzi (Marsupialia : Macropodidae) in captivity, with age estimation and development of the pouch young. Wildl. Res. 30, 505. https://doi.org/10.1071/WR02090

**Klingenberg**, C.P., 2016. Size, shape, and form: concepts of allometry in geometric morphometrics. Development genes and evolution 226, 113–137.

**Köhler**, M., Herridge, V., Nacarino-Meneses, C., Fortuny, J., Moncunill-Solé, B., Rosso, A., Sanfilippo, R., Palombo, M.R., Moyà-Solà, S., 2021. Palaeohistology reveals a slow pace of life for the dwarfed Sicilian elephant. Sci Rep 11, 22862. https://doi.org/10.1038/s41598-021-02192-4

**Larramendi**, A., 2016. Proboscideans: Shoulder Height, Body Mass and Shape. APP. https://doi.org/10.4202/app.00136.2014

**Larramendi**, A., Palombo, M.R., 2015. Body Size, Structure, Biology and Encephalization Quotient of Palaeoloxodon ex gr. P. falconeri from Spinagallo Cave (Hyblean plateau, Sicily). Hystrix, the Italian Journal of Mammalogy 26. https://doi.org/10.4404/hystrix-26.2-11478

**Lieberman**, D.E., Carlo, J., Ponce De León, M., Zollikofer, C.P.E., 2007. A geometric morphometric analysis of heterochrony in the cranium of chimpanzees and bonobos. Journal of Human Evolution 52, 647–662. https://doi.org/10.1016/j.jhevol.2006.12.005

**Lister**, A., 1996. Dwarfing in island elephants and deer: Processes in relation to time of isolation. Symposia of the Zoological Society of London 69, 277–292.

**Lister**, A.M., 1989. Rapid dwarfing of red deer on Jersey in the Last Interglacial. Nature 342, 539–542. https://doi.org/10.1038/342539a0

**Liu**, A.G.S.C., Seiffert, E.R., Simons, E.L., 2008. Stable isotope evidence for an amphibious phase in early proboscidean evolution. Proc. Natl. Acad. Sci. U.S.A. 105, 5786–5791. https://doi.org/10.1073/pnas.0800884105

**Lomolino**, M.V., 1985. Body Size of Mammals on Islands: The Island Rule Reexamined. The American Naturalist 125, 310–316. https://doi.org/10.1086/284343

**Mahboubi**, S., Bocherens, H., Scheffler, M., Benammi, M., Jaeger, J.-J., 2014. Was the Early Eocene proboscidean Numidotherium koholense semi-aquatic or terrestrial? Evidence from stable isotopes and bone histology. Comptes Rendus Palevol 13, 501–509. https://doi.org/10.1016/j.crpv.2014.01.002

**Meyer**, M., Palkopoulou, E., Baleka, S., Stiller, M., Penkman, K. E., Alt, K. W., ... & Hofreiter, M. 2017. Palaeogenomes of Eurasian straight-tusked elephants challenge the current view of elephant evolution. Elife, 6, e25413 **Mitsopoulou**, V., Michailidis, D., Theodorou, E., Isidorou, S., Roussiakis, S., Vasilopoulos, T., Polydoras, S., Kaisarlis, G., Spitas, V., Stathopoulou, E., Provatidis, C., Theodorou, G., 2015. Digitizing, modelling and 3D printing of skeletal digital models of Palaeoloxodon tiliensis (Tilos, Dodecanese, Greece). Quaternary International 379, 4–13. https://doi.org/10.1016/j.quaint.2015.06.068

**Nganvongpanit**, K., Siengdee, P., Buddhachat, K., Brown, J.L., Klinhom, S., Pitakarnnop, T., Angkawanish, T., Thitaram, C., 2017. Anatomy, histology and elemental profile of long bones and ribs of the Asian elephant (Elephas maximus). Anat Sci Int 92, 554–568. https://doi.org/10.1007/s12565-016-0361-y

**Nomikou**, P., Papanikolaou, D., 2017. A comparative morphological study of the kos-nisyros-tilos volcanosedimentary basins. geosociety 43, 464. https://doi.org/10.12681/bgsg.11197

**Osborn**, H.F., 1936. Proboscidea: Moeritherioidea, Deinotherioidea, Mastodontoidea. American Museum Press.

**Osborn**, H.F., 1929. The titanotheres of ancient Wyoming, Dakota, and Nebraska. Department of the Interior, US Geological Survey.

**Osborn**, H.F., 1942. Proboscidea, Vol. II. The American Museum of Natural History Press, New York, 805-1675.

**Palombo**, M., 2001. Paedomorphic features and allometric growth in the skull of Elephas falconeri from Spinagallo (Middle Pleistocene, Sicily), in: The World of Elephants. Proceedings of the First International Congress. Consiglio Nazionale Delle Richerche, Rome. pp. 492–496.

**Palombo**, M.R., 2003. Elephas? Mammuthus? Loxodonta? The question of the true ancestor of the smallest dwarfed elephant of Sicily. Deinsea 9, 273–292.

**Palombo**, M.R., Ferretti, M.P., Pillola, G.L., Chiappini, L., 2012. A reappraisal of the dwarfed mammoth Mammuthus lamarmorai () from Gonnesa (south-western Sardinia, Italy). Quaternary International 255, 158–170. https://doi.org/10.1016/j.quaint.2011.05.037

**Palombo**, M.R., Moncunill-Solé, B., 2025. Dwarfing and gigantism in Quaternary vertebrates, in: Encyclopedia of Quaternary Science, Elsevier, 584-608. DOI: 10.1016/B978-0-323-99931-1.00012-X

**Palombo**, M.R., Zedda, M., Melis, R.T., 2017. A new elephant fossil from the late Pleistocene of Alghero: The puzzling question of Sardinian dwarf elephants. Comptes Rendus Palevol 16, 841–849. https://doi.org/10.1016/j.crpv.2017.05.007

**Prothero**, D.R., Sereno, P.C., 1982. Allometry and Paleoecology of Medial Miocene Dwarf Rhinoceroses from the Texas Gulf Coastal Plain. Paleobiology 8, 16–30. https://doi.org/10.1017/S0094837300004322

**R Core Team** (2021). R: A language and environment for statistical computing. R Foundation for Statistical Computing, Vienna, Austria. URL https://www.R-project.org/

**Rohlf**, F.J., Slice, D., 1990. Extensions of the Procrustes method for the optimal superimposition of landmarks. Systematic Biology 39, 40–59.

**Roth**, V., Damuth, J., MacFadden, B., 1990. Insular dwarf elephants: a case study in body mass estimation and ecological inference. Body size in mammalian paleobiology: Estimation and biological implications 151–179.

**Roth**, V.L., 1984. How elephants grow: heterochrony and the calibration of developmental stages in some living and fossil species. Journal of Vertebrate Paleontology 4, 126–145. https://doi.org/10.1080/02724634.1984.10011993

**Sanders,** M., P., Mateus, O., Laven, T., Knötschke, N., 2006. Bone histology indicates insular dwarfism in a new Late Jurassic sauropod dinosaur. Nature 441, 739–741. https://doi.org/10.1038/nature04633

**Sanders,** W.J. 2024. Evolution and fossil record of African Proboscidea. CRC Press, Taylor and Francis Group, LLC. 346 pp. doi:10.1201/b20016 **Scarborough**, M.E., 2022. Extreme body size variation in Pleistocene dwarf elephants from the Siculo-Maltese palaeoarchipelago: disentangling the causes in time and space. Quaternary 5, 17.

**Scarborough**, M.E., 2020. Insular adaptations in the appendicular skeleton of Sicilian and Maltese dwarf elephants. PhD thesis. Faculty of Science, Department of Biological Sciences, University of Cape Town, 292 pp. http://hdl.handle.net/11427/32747.

**Schlager**, S., 2017. Morpho and Rvcg–Shape Analysis in R: R-Packages for geometric morphometrics, shape analysis and surface manipulations, in: Zheng, G., Li, S., Szekely, G. (Eds.), Statistical Shape and Deformation Analysis. Academic Press, Cambridge, pp. 217–256.

**Shindo**, T., Mori, M., 1956. Musculature of Indian Elephant. Part 1. Musculature of the Forelimb. Okajimas Folia Anatomica Japonica 28, 89–113. https://doi.org/10.2535/ofaj1936.28.1-6_89

**Shoshani**, J., Tassy, P., 2005. Advances in proboscidean taxonomy & classification, anatomy & physiology, and ecology & behavior. Quaternary International 126, 5–20.

**Smuts**, M.M., Bezuidenhout, A.J., 1994. Osteology of the pelvic limb of the African elephant (Loxodonta africana). Onderstepoort J Vet Res 61, 51–66.

**Smuts**, M.M., Bezuidenhout, A.J., 1993. Osteology of the thoracic limb of the African elephant (Loxodonta africana). Onderstepoort J Vet Res 60, 1–14.

**Sondaar**, P.Y., 1977. Insularity and Its Effect on Mammal Evolution, in: Hecht, M.K., Goody, P.C., Hecht, B.M. (Eds.), Major Patterns in Vertebrate Evolution. Springer US, Boston, MA, pp. 671–707. https://doi.org/10.1007/978-1-4684-8851-7_23

**Theodorou**, G., Symeonidis, N., Stathopoulou, E., 2007. Elephas tiliensis n. sp. from Tilos island (Dodecanese, Greece). Hellenic Journal of Geosciences 42, 19–32.

**Trenkwalder**, H., 2013. Über die Muskulatur des Schultergürtels und der proximalen Vorderextremität des afrikanischen Elefanten (Loxodonta africana). PhD thesis.

Veterinärmedizinische Universität Wien, Institut für Anatomie, Histologie und Embryologie. 131p.

**Valen**, L.V., 1973. Body Size and Numbers of Plants and Animals. Evolution 27, 27. https://doi.org/10.2307/2407116

**van den Bergh**, G.D., 1999. The Late Neogene elephantoid-bearing faunas of Indonesia and their palaeozoogeographic implications . Scripta Geologica 117, 1–419.

**van den Bergh**, G.D., Awe, R.D., Morwood, M.J., Sutikna, T., Jatmiko, Wahyu Saptomo, E., 2008. The youngest stegodon remains in Southeast Asia from the Late Pleistocene archaeological site Liang Bua, Flores, Indonesia. Quaternary International 182, 16–48. https://doi.org/10.1016/j.quaint.2007.02.001

**van der Geer**, A., 2005. Island ruminants and parallel evolution of functional structures. Quaternaire 2, 231–240.

**van der Geer**, A.A., Lyras, G.A., Lomolino, M.V., Palombo, M.R., Sax, D.F., 2013. Body size evolution of palaeo-insular mammals: temporal variations and interspecific interactions. Journal of Biogeography 40, 1440–1450. https://doi.org/10.1111/jbi.12119

**van der Geer**, A.A.E., 2014. Parallel patterns and trends in functional structures in extinct island mammals. Integrative Zoology 9, 167–182. https://doi.org/10.1111/1749-4877.12066

**van der Geer**, A.A.E., Lyras, G.A., Mitteroecker, P., MacPhee, R.D.E., 2018. From Jumbo to Dumbo: Cranial Shape Changes in Elephants and Hippos During Phyletic Dwarfing. Evol Biol 45, 303–317. https://doi.org/10.1007/s11692-018-9451-1

**van der Geer**, A.A.E., Van Den Bergh, G.D., Lyras, G.A., Prasetyo, U.W., Due, R.A., Setiyabudi, E., Drinia, H., 2016. The effect of area and isolation on insular dwarf proboscideans. Journal of Biogeography 43, 1656–1666. https://doi.org/10.1111/jbi.12743

**Wibowo**, U.P., 2016. Walking with Indonesian elephants: attribution of isolated proboscidean femurs and tibias to genus based on morphological differences. PhD thesis. University of Wollongong. 200p.

**Wiley**, D.F., Amenta, N., Alcantara, D.A., Ghosh, D., Kil, Y.J., Delson, E., Harcourt-Smith, W., Rohlf, F.J., St John, K., Hamann, B., 2005. Evolutionary morphing, in: Proceedings of IEEE Visualization 2005. Presented at the Proceedings of IEEE visualization 2005, IEEE, Piscataway, pp. 431–438.

# Figure legends

**Figure 1**: (A) Cladogram of the sample studied based in part on ancient DNA genomics (Meyer et al. 2017). Dwarf elephants are in bold. Modified from Baleka et al. (2022); (B) Acquisition methods for the associated analyses.

**Figure 2**: Virtual slices of the humeri of (A), (B), (E), (G), H) an adult *Palaeoloxodon tiliensis* specimen (NHMW-Geo-1976/1833/0007) and (C), (D), (F) a *Palaeoloxodon falconeri* fetus/neonate (NMB-Ty.12560) in (A), (C), (G), (H), (K) coronal, (B), (D) sagittal, and (E), (F) transversal view. The red lines show the direction of trabeculae in highly anisotropic areas. D.c., deltoid crest, G.t., greater trochanter, L.e., lateral epicondyle, M.e., medial epicondyle, O.f., olecranon fossa, S.c., supracondylar crest. Cran, cranial, Lat, lateral, Med, medial, Prox, proximal.

**Figure 3**: Virtual slices of the radii of (A), (B), (E), (G), (H) an adult *Palaeoloxodon tiliensis* specimen (NHMW-Geo-1976/1833/0006) and (C), (D), (F) a *Palaeoloxodon falconeri* fetus/neonate (NMB-Ty.12561) in (A), (C), (G) coronal, (B), (D), (H), sagittal, (E), (F) transversal view. The red lines show the direction of trabeculae in highly

anisotropic areas. S.c.h., surface of contact with the humerus, S.d.u., distal surface of contact with the ulna, S.p.u., proximal surface of contact with the ulna. Caud, caudal, Cran, cranial, Lat, lateral, Med, medial, Prox, proximal.

**Figure 4**: Virtual slices of the ulnae of (A), (B), (C), (F), (G), (H) an adult *Palaeoloxodon tiliensis* specimen (NHMW-Geo-1976/1833/0008) and (D), (E), (I) a *Palaeoloxodon falconeri* fetus/neonate (NMB-Ty.12561) in (A), (D), (F), (G) sagittal, (B), (C), (E) coronal, (H), (I) transversal view. The red lines show the direction of trabeculae in highly anisotropic areas. A.f.c, articular facet for the carpus, A.p., anconeal process, L.c.p., lateral coronoid process, M.c.p., medial coronoid process, O., olecranon, R.d.e., radial distal epiphysis, S.d.r., distal surface of contact with the radius, S.p., styloid process, T.n., trochlear notch, U.d.e., ulnar distal epiphysis. Cran, cranial, Lat, lateral, Med, medial, Prox, proximal.

**Figure 5**: Virtual slices of the femora of (A), (B), (G), (J), (K) an adult *Palaeoloxodon tiliensis* specimen (NHMW-Geo-1976/1833/0001), and (C), (D), (E), (F), (H), (I) *Palaeoloxodon falconeri* fetus/neonate and juvenile: (C), (D), (H), NMB-Ty.12557, (E), (F), (I), NMB-Ty.12952 in (A), (C), (E), (J), (K) coronal, (B), (D), (F) sagittal, (G), (H), (I) transversal view. The red lines show the direction of trabeculae in highly anisotropic areas. G.t., greater trochanter, L.c., lateral condyle, L.e., lateral epicondyle, L.t., lesser trochanter, M.c., medial condyle, M.e., medial epicondyle. Cran, cranial, Lat, lateral, Med, medial, Prox, proximal.

**Figure 6**: Virtual slices of the tibiae of adult *Palaeoloxodon tiliensis* specimens, (A), (B), (E), (G), (I) NHMW-Geo-1976/1833/0003, (C), (D), (F), (H) NHMW-Geo-1976/1833/0002 in (A), (C), (H), (I), coronal, (B), (D), (G) sagittal, (E), (F) transversal view. The red lines show the direction of trabeculae in highly anisotropic areas. C.c., cranial crest, C.o., cochlea, I.e., intercondylar eminence, L.c., lateral condyle, M.c., medial condyle, S.c.f., surface of contact with the fibula, T.m., tibial malleolus. Caud, caudal, Lat, lateral, Med, medial, Prox, proximal.

**Figure 7**: Virtual slices of the fibulae of an adult *Palaeoloxodon tiliensis* specimens (NHMW-Geo-1976/1833/0004), in (A), (B), (F) sagittal, (C), (D) coronal and (E) transversal view. The red lines show the direction of trabeculae in highly anisotropic areas. A.f.m., articular facet of the malleolus, A.t.c., articular facet for the talus and the calcaneum. Cran, cranial, Lat, lateral, Med, medial, Prox, proximal.

**Figure 8**: Visualizations of the cortical thickness mapping of the six limb long bones of adult specimens of *Palaeoloxodon tiliensis* and *Loxodonta africana* in (a) cranial, (b) lateral, (c) caudal and (d) medial view. Cortical

thickness is represented by a gradient ranging from cold (low cortical thickness) to warm (high cortical thickness) colors on the 3D mapping; values are relative to the minimum and maximum cortical thickness for each bone.

**Figure 9**: Visualizations of the cortical thickness mapping of the six limb long bones of juvenile specimens of *Palaeoloxodon falconeri* and *Loxodonta africana* in (a) cranial, (b) lateral, (c) caudal and (d) medial view. Cortical thickness is represented by a gradient ranging from cold (low cortical thickness) to warm (high cortical thickness) colors on the 3D mapping; values are relative to the minimum and maximum cortical thickness for each bone.

**Figure 10**: Visualizations of the mean shape the humerus of (A), (B) juveniles and (C), (D) adult specimens of (A), (C), *Elephas maximus* and (B), (D) *Loxodonta africana* specimens in (a) cranial, (b) lateral, (c) caudal and (d) medial views.

**Figure 11**: Visualizations of the mean shape the radius and ulna of (A) juvenile and (B) adult specimens of *Elephas maximus* in (a) cranial, (b) lateral, (c) caudal and (d) medial views.

**Figure 12**: Visualizations of the limb long bones of an adult *Palaeoloxodon antiquus*, an adult *Palaeoloxodon tiliensis* and a juvenile *Loxodonta africana* with features of interest annotated. (A) Humerus, (B) radius, (C) ulna, (D) femur, (E) tibia and (F) fibula. A.s.c., articular surface for the carpal bones, A.s.f., articular surface for the fibula, D.t., deltoid crest, Fi.h., fibular head, F.h., femoral head, F.n., femoral neck, Fi.n., fibular notch, F.t., femoral trochlea, G.t., greater trochanter, H.t., humeral trochlea, I.e., intercondylar eminence, I.g., intertubercular groove, L.c., lateral condyle, L.t., lesser trochanter, O.t., olecranon tuberosity, R.h., radial head, T.n., trochlear notch. Cran, cranial, Med, medial, Lat, lateral, Prox, proximal.

**Figure 13**: (A) Results of the PCA performed on shape data of the humerus of all specimens along with the visualizations of the theoretical shapes at the minimum and maximum of the first two axes; The size of the points is proportional to the centroid size of the bones. (B) Results of the Neighbour-Joining tree computed on Euclidean distances between bone shapes.

**Figure 14**: (A) Results of the PCA performed on shape data of the ulna of all specimens along with the visualizations of the theoretical shapes at the minimum and maximum of the first two axes; The size of the points is proportional to the centroid size of the bones. (B) Results of the Neighbour-Joining tree computed on Euclidean distances between bone shapes.

**Figure 15**: (A) Results of the PCA performed on shape data of the femur of all specimens along with the visualizations of the theoretical shapes at the minimum and maximum of the first two axes; The size of the points is proportional to the centroid size of the bones. (B) Results of the Neighbour-Joining tree computed on Euclidean distances between bone shapes.

**Figure 16**: (A) Results of the PCA performed on shape data of the tibia of all specimens along with the visualizations of the theoretical shapes at the minimum and maximum of the first two axes; The size of the points is proportional to the centroid size of the bones. (B) Results of the Neighbour-Joining tree computed on Euclidean distances between bone shapes.

**Figure 17**: (A) Results of the PCA performed on shape data of the fibula of all specimens along with the visualizations of the theoretical shapes at the minimum and maximum of the first two axes; The size of the points is proportional to the centroid size of the bones. (B) Results of the Neighbour-Joining tree computed on Euclidean distances between bone shapes.

**Figure 18**: Boxplots of the robustness of each bone according to genus and age. *$p<0.05$, **$p<0.01$, ***$p<0.001$

**Figure 19:** Morphological and microanatomical traits in the limb long bones of proboscideans species sampled. Each morphological character has been discretised into 2 to 3 states according to intensity, ranging from state 1 to state 3.

**Table 1:** Results of the Procrustes ANOVAs testing for correlation between shape data and log-transformed centroid size among (1) all specimens, (2) non-dwarf specimens, (3) adult specimens and (4) adult non-dwarf specimens. p, p-value, $r^2$, determination coefficient value. Significant results are in bold.

**Table 2**: Results of the correlation tests between the centroid size and the different size parameters for each bone. *Ci*, smallest diaphyseal circumference; *MaxL*, maximum length of the bone, *Rb*, robustness; NA, not available since we could not measure the circumference in the radius and ulna; p, p-value; $r^2$, determination coefficient value. Significant results are in bold.

# Supplementary material legends

**Supplementary Table 1:** Sample studied. OS, ontogenetic stage; A, adult, C, calf, F, fetus, J, juvenile, NE, neonate, S, subadult; GMMs, Geometric Morphometrics analyses; Y, yes; N, no. Institutional codes: AMNH, American Museum of Natural History, New-York (USA); ANSP, Academy of Natural Sciences of Drexel University, Philadelphia (USA); CCEC, Centre de Conservation et d'Étude des Collections, Musée des Confluences, Lyon (France); HNHM, Hungarian Natural History Museum, Budapest (Hungary); IMNH, Idaho Museum of Natural History, Pocatello (USA); ISEM, Institut des Sciences de l'Évolution, Université de Montpellier, Montpellier (France); MAFI, Magyar Állami Földtani Intézet, Budapest (Hungary); MGP, Museo di Geologia e Paleontologia, Università degli studi di Firenze, Florence (Italy); MNHN, Muséum national d'Histoire naturelle, Paris (France); NBC, Naturalis Biodiversity Center, Leiden (Netherlands); NHMUK, Natural History Museum, London (UK); ); NHMW, Naturhistorisches Museum Wien, Vienna (Austria); NMB, Naturhistorisches Museum Basel, Basel (Switzerland); RBINS, Royal Belgian Institute of Natural Sciences, Brussels (Belgium); RMCA, Royal Museum for Central Africa, Tervuren (Belgium); RVC, Royal Veterinary College, London (UK); SMNS, Staatliches Museum für Naturkunde Stuttgart, Stuttgart (Germany); UF, University of Florida, Gainesville (USA); ZSM, Zoologische Staatssammlung München, Munich (Germany). Abbreviations: Fe, femur; Fi, fibula; H, humerus; na, not available; R, radius; T, tibia; U, ulna.

**Supplementary Table 2:** Results of Procrustes ANOVAs testing for shape difference between adult and juvenile specimens among the E. maximus and L. africana samples. p, p-value, $r^2$, determination coefficient value. Significant results are in bold.

**Supplementary Table 3**: Results of the ANOVAs testing for robustness variation between genera and between ontological stages. Rb, robustness.

**Supplementary Table 4**: Results of the correlation tests between the size parameters and the two first principal components of the principal components analyses computed using the shape data of the entire sample for each bone. Ci, smallest diaphyseal circumference; Cs, centroid size; MaxL, maximum length of the bone; NA, not available since we could not measure the circumference in the radius and ulna; p, p-value; $r^2$, determination coefficient value. Significant results are in bold.

**Supplementary Figure 1**: Allometric trends in the (A) humerus, (B) radius, (C) ulna, (D) femur, (E) tibia, (F) fibula